# Density and excess molar enthalpy of 2-butanol + glyme liquid mixtures. Application of the DISQUAC, UNIFAC (Dortmund) and ERAS models to 2-alkanol + linear ether systems

Fatemeh Pazoki, João Victor Alves-Laurentino, Susana Villa, Daniel Lozano-Martín, Fernando Hevia*, Luis Felipe Sanz, Juan Antonio González

GETEF. Departamento de Física Aplicada. Facultad de Ciencias. Universidad de Valladolid. Paseo de Belén, 7, 47011 Valladolid, Spain.

* Corresponding author, e-mail: luisfernando.hevia@uva.es

## Abstract

Densities and excess molar volumes ($V_m^E$) at (293.15 to 303.15) K and excess molar enthalpies ($H_m^E$) at 298.15 K are reported for the 2-butanol + 2,5-dioxahexane, or + 2,5,8-trioxanonane, or + 2,5,8,11-tetraoxadodecane, or + 2,5,8,11,14-pentaoxapentadecane systems at 95 kPa. Densities were measured using a densitometer Anton Paar DSA 5000, and excess enthalpies by means of a micro-calorimeter Tian-Calvet. $H_m^E$ and $V_m^E$ data are positive, indicating that interactions between like molecules are dominant and that the main contribution to $V_m^E$ is the interactional one. Our enthalpic data together with data from the literature have been used to evaluate the enthalpy of alcohol-glyme interactions, which are ranged between (−31 and −22) kJ mol$^{-1}$. From the comparison with $H_m^E$ results for glyme + heptane systems, and from the position of the investigated mixtures in the diagram $G_m^E$ (excess molar Gibbs energy) *vs.* $H_m^E$, it is shown that interactions are of dispersive type. 2-Alkanol + dipropyl ether (DPE), or + glyme systems have also been studied using the DISQUAC, UNIFAC (Dortmund) and ERAS models. According to ERAS results, physical interactions in glyme systems are dominant, while association effects are predominant in mixtures with DPE. The DISQUAC interaction parameters suggest that interactions related to the OH/O contacts are essentially dispersive. UNIFAC fails when representing thermodynamic properties of the studied solutions.

# 1. Introduction

This work is part of experimental and theoretical research on liquid mixtures which contain the O and/or the OH groups in the same or in different molecules. In the past, we have provided data on liquid-liquid equilibria for linear polyether + alkane mixtures [1], and excess molar volumes ($V_{\mathrm{m}}^{\mathrm{E}}$) [2–8] and enthalpies ($H_{\mathrm{m}}^{\mathrm{E}}$) [9–11] for 1-alkanol + linear ether systems. The complexity of the latter solutions arises from the partial destruction of the alcohol network by the active ether molecules, and from the new OH---O bonds created upon mixing [12]. For systems with 1-alkanols longer than ethanol, this destruction of the alcohol network is particularly crucial when linear polyethers are involved, since they lead to weak orientational effects in the corresponding solutions [7]. We have also investigated proximity effects between the O and OH groups within the same molecule by means of systems containing hydroxyethers [10,13–18]. The theoretical research [6,7,19–22] has been carried out using different theories such as DISQUAC [23], ERAS [24], Flory [25] or the Kirkwood-Buff integrals formalism [26]. More recently, and to explore the influence of the steric hindrance of the OH group in secondary alcohols on the thermodynamic properties of their solutions, we have started systematic research on mixtures containing 2-alkanols [27,28]. As a continuation of a preceding work on 2-propanol (2PrOH) + glyme mixtures [28], we provide now experimental data on densities and $H_{\mathrm{m}}^{\mathrm{E}}$ for the systems 2-butanol (2BuOH) + 2,5-dioxahexane, or + 2,5,8-trioxanonane, or 2,5,8,11-tetraoxadodecane, or + 2,5,8,11,14-pentaoxapentadecane at 95 kPa. From the densities measured at (293.15-303.15) K, we also provide $V_{\mathrm{m}}^{\mathrm{E}}$ data for the same temperatures. The $H_{\mathrm{m}}^{\mathrm{E}}$ values are given at 298.15 K. Along the manuscript, the mentioned glymes are represented by their number of O atoms, $u$ (*e.g,* $u$ = 3 stands for 2,5,8-trioxanonane; the value $u$ = 1 is freely assigned to dipropyl ether (DPE)). The mixtures are also studied using three different models: DISQUAC, UNIFAC (Dortmund) [29] and ERAS. UNIFAC calculations were conducted using interaction parameters from the literature [30].

Alcohol + ether mixtures have a wide variety of applications. They are used as gasoline additives due to their octane-enhancing and pollution reducing properties [31,32] or as working fluids for absorption refrigerant machines to improve the cycle machine [33]. Systems formed by a short-chain 1-alkanol and a linear polyether are used as simple models of complex systems water + polyethylene glycols, which are very common in biochemical and biomedical processes [34,35]. In addition, alkanol + ether mixtures are industrially relevant since alkanols are basic components in the synthesis of oxaalkanes and therefore are contained as an impurity.

# 2. Materials and methods

## 2.1. Materials

Pure liquids were used as received from the supplier, without any further purification. The source and purity of the chemicals are shown in Table 1. The composition of the mixtures (given by the mole fraction of 2BuOH, $x_1$) was calculated from mass measurements using an analytical balance (MSU125P, Sartorius) after correcting buoyancy effects. The standard uncertainty of these measurements is $5\times10^{-5}$ g. Molar quantities were calculated using the relative atomic mass Table of 2015 issued by I.U.P.A.C [36]. The standard uncertainty in the final mole fraction is estimated to be 0.0005. All measurements were performed at 95 kPa. The temperature (standard uncertainty of 0.01 K) of the samples was measured by means of Pt-100 resistances, calibrated using the triple point of water and the melting point of Ga.

**Table 1**. Sample description.

| Chemical name | CAS Number | Source | [a] Initial purity | Purification method |
|---|---|---|---|---|
| 2-butanol | 78-92-2 | Sigma-Aldrich | 0.997 | none |
| 2,5-dioxahexane (monoglyme) | 110-71-4 | Sigma-Aldrich | 0.999 | none |
| 2,5,8-trioxanonane (diglyme) | 111-96-6 | Fluka | 0.998 | none |
| 2,5,8,11-tetraoxadodecane (triglyme) | 112-49-2 | Sigma-Aldrich | 0.999 | none |
| 2,5,8,11,14-pentaoxapentadecane (tetraglyme) | 143-24-8 | Sigma-Aldrich | 0.999 | none |

[a] Gas chromatography area fraction. Certified by the supplier.

## 2.2. Density measurements

Densities of the liquid samples were measured by means of a densitometer Anton Paar DSA5000. Its temperature stability is 0.001 K, with control provided by Peltier modules.

For this kind of measurements, the standard uncertainty under repeatability conditions is estimated to be $5\times10^{-5}$ g cm$^{-3}$, which provides criteria for the number of significant figures. Several measurements of the densities ($\rho$) of the pure liquids were conducted over time to test their stability, remaining constant within the measurement uncertainty. The apparatus was calibrated using as reference pure liquids: heptane, isooctane, cyclohexane, toluene and water HPLC plus. The procedure was applied at each of the three considered temperatures. More details on the calibration and on the applied procedure can be encountered elsewhere [28,37,38]. A comparison between our $\rho$ values at (293.15-303.15) K with literature results is shown in Table 2. The maximum differences are of order 0.1 %. The total relative expanded uncertainty (with a coverage factor of 2) can be then estimated equal to 0.002. The densitometer was also tested by measuring $V_{\mathrm{m}}^{\mathrm{E}}$ for the reference system cyclohexane + benzene [39]. Although the differences between literature values and our results are lower than 1 %, the standard uncertainty for $V_{\mathrm{m}}^{\mathrm{E}}$ is estimated to be $0.010\left|V_{\mathrm{m}}^{\mathrm{E}}\right|_{\max}$ + 0.005 cm$^3$ mol$^{-1}$ ($\left|V_{\mathrm{m}}^{\mathrm{E}}\right|_{\max}$ denotes the maximum experimental value of $\left|V_{\mathrm{m}}^{\mathrm{E}}\right|$ with respect to the mole fraction). Table 2 also lists the isobaric thermal expansion coefficients, $\alpha_p$, at $T$ = 298.15 K of the pure compounds obtained from the equation:

$$\alpha_p = -\frac{1}{\rho}\left(\frac{\partial\rho}{\partial T}\right)_p \quad (1)$$

where $\rho$ is the density at $T$ = 298.15 K and $(\partial\rho/\partial T)_p$ is calculated as the slope of the linear regression of $\rho$ as a function of $T$ in the range $T$ = (293.15 to 303.15) K. The comparison of our values with literature results is good. The relative standard uncertainty of the $\alpha_p$ values is 0.02. For later calculations, the literature values of isothermal compressibilities, $\kappa_T$, of the pure compounds at $T$ = 298.15 K are given in Table S1 (supplementary material).

**Table 2.** Density, $\rho$, and isobaric thermal expansion coefficient, $\alpha_p$, of the pure liquids used in this work at temperature $T$ and pressure $p$ = 95 kPa [a]. $\sigma(F)$, root-mean-square deviation of the literature values from the experimental values at temperature $T$ = 298.15 K, equation (6).

| Compound | Property | $T$/K | Exp. | Lit. |
|---|---|---|---|---|
| 2-butanol | $\rho/(\text{g cm}^{-3})$ | 293.15 | 0.80640 | 0.8067 [40], 0.8067 [41], 0.8062 [42], 0.80648 [43], 0.8067 [44] |
| | | 298.15 | 0.80228 | 0.8023 [45], 0.8026 [40], 0.80241 [46], 0.80219 [47], 0.802528 [48], 0.8027 [41], 0.8020 [42], 0.80235 [43], 0.8027 [44] |
| | | 303.15 | 0.79809 | 0.7984 [41], 0.7979 [42], 0.7984 [44] |
| | $\alpha_p/(10^{-3}\ \text{K}^{-1})$ | 298.15 | 1.036 | 1.022[b] [45], 1.05 [40], 1.034[b] [46], 1.035[b] [47], 1.034[b] [41], 1.035[b] [42], 1.034[b] [44] |
| monoglyme | $\rho/(\text{g cm}^{-3})$ | 293.15 | 0.86724 | 0.86671 [49], 0.86765 [50], 0.8670 [51], 0.8667 [52] |
| | | 298.15 | 0.86177 | 0.86183 [53], 0.86114 [54], 0.86114 [55], 0.86110 [56], 0.8612 [49], 0.9615 [51], 0.8611 [52], 0.8616 [57], 0.86208 [58] |
| | | 303.15 | 0.85628 | 0.8557 [49], 0.85718 [50], 0.8557 [51], 0.8555 [52] |
| | $\alpha_p/(10^{-3}\ \text{K}^{-1})$ | 298.15 | 1.272 | 1.268 [53], 1.277[b] [54], 1.274[b] [55], 1.290[b] [56], 1.278[b] [49] |
| diglyme | $\rho/(\text{g cm}^{-3})$ | 293.15 | 0.94395 | 0.94511 [50] ,0.9434 [51] ,0.9435 [52] ,0.94405 [59], 0.9452 [60] |
| | | 298.15 | 0.93899 | 0.93872 [53], 0.93935 [54], 0.93883 [55], 0.93859 [56], 0.9385 [51], 0.9386 [52], 0.93906 [59], 0.9402 [60] |
| | | 303.15 | 0.93402 | 0.93569 [50], 0.9335 [51], 0.9337 [52], 0.93407 [59], 0.9352 [60] |
| | $\alpha_p/(10^{-3}\ \text{K}^{-1})$ | 298.15 | 1.058 | 1.060 [53], 1.064[b] [54], 1.057[b] [55], 1.055[b] [56], 1.055[b] [51], 1.063[b] [59], 1.064[b] [60] |
| triglyme | $\rho/(\text{g cm}^{-3})$ | 293.15 | 0.98491 | 0.98569 [50], 0.9853 [61], 0.9860 [60] |
| | | 298.15 | 0.98016 | 0.98079 [53], 0.98058 [54], 0.98049 [62], 0.9805 [61], 0.98064 [63], 0.98001 [64], 0.9815 [60] |
| | | 303.15 | 0.97541 | 0.97672 [50], 0.9767 [60] |
| | $\alpha_p/(10^{-3}\ \text{K}^{-1})$ | 298.15 | 0.969 | 0.965 [53], 0.969[b] [54], 0.969[b] [62], 0.966[b] [61], 0.969[b] [63], 0.948[b] [60] |
| tetraglyme | $\rho/(\text{g cm}^{-3})$ | 293.15 | 1.01090 | 1.0109 [65], 1.01160 [50], 1.0123 [60], 1.0114 [66] |
| | | 298.15 | 1.00626 | 1.00569 [53], 1.00668 [54], 1.00716 [67], 1.00568 [68], 1.00662 [64], 1.0076 [60] |
| | | 303.15 | 1.00163 | 1.0017 [65], 1.00290 [50], 1.0031 [60] |
| | $\alpha_p/(10^{-3}\ \text{K}^{-1})$ | 298.15 | 0.921 | 0.921 [53], 0.928[b] [54], 0.915[b] [67], 0.929[b] [68], 0.913[b] [60] |

[a] Expanded uncertainties, $U$, and relative expanded uncertainties, $U_r$, with a coverage factor of 2: $U(T)$ = 0.02 K, $U(p)$ = 10 kPa, $U_r(\rho)$ = 0.002, $U_r(\alpha_p)$ = 0.04; [b] calculated from data reported in the reference.

## 2.3. Excess molar enthalpy measurements

This type of measurement was carried out at 298.15 K using a self-manufactured standard Tian-Calvet micro-calorimeter. Its temperature stability is 0.01 K, with control provided by PID-assisted heating resistors. During the measurements, the laboratory temperature was maintained below the measuring temperature. The mixing cell, also designed by the authors, is made of aluminum and has a small gas phase (< 2 %). More details on the mixing process and on the calibration of the apparatus can be found elsewhere [28,69]. We merely underline that the apparatus was calibrated electrically at 298.15 K and tested performing measurements for the benzene + cyclohexane mixture [70,71]. The observed deviations are ≤ 1.2 % and the estimated maximum relative expanded uncertainty (with a coverage factor of 2) for $H_\mathrm{m}^\mathrm{E}$ is 0.03.

# 3. Experimental results

The experimental $\rho$ and $V_\mathrm{m}^\mathrm{E}$ values of 2-BuOH + glyme mixtures at $p$ = 95 kPa and $T$ = (293.15 to 303.15) K are included in Table 3 as functions of $x_1$. Values of $A_p = \left(\partial V_\mathrm{m}^\mathrm{E}/\partial T\right)_{p,x}$ at $p$ = 95 kPa and $T$ = 298.15 K, calculated as the slope of linear regressions of $V_\mathrm{m}^\mathrm{E}$ as a function of $T$ in the range of the measured temperatures, are also listed in Table 3. The experimental values of $H_\mathrm{m}^\mathrm{E}$ at 95 kPa and $T$ = 298.15 K are collected in Table 4. From the $V_\mathrm{m}^\mathrm{E}$ and $H_\mathrm{m}^\mathrm{E}$ data, we have also determined the isochoric excess molar internal energies, $U_{V\mathrm{m}}^\mathrm{E}$, using the equation [72]:

$$U_{V\mathrm{m}}^\mathrm{E} = H_\mathrm{m}^\mathrm{E} - T\frac{\alpha_p}{\kappa_T}V_\mathrm{m}^\mathrm{E} \qquad (2)$$

In equation (2), $T\left(\alpha_p/\kappa_T\right)V_\mathrm{m}^\mathrm{E}$ is called the equation-of-state term (EoS term) and is related to the volume variation during mixing. Along with calculations, and as in other previous applications, the values of $\alpha_p$ and $\kappa_T$ of the mixtures were determined by assuming ideal behavior for these quantities [73,74].

The excess quantities are fitted to Redlich-Kister polynomials [75]:

$$F_\mathrm{m}^\mathrm{E} = x_1(1-x_1)\sum_{i=0}^{k-1} A_i(2x_1-1)^i \qquad (F_\mathrm{m}^\mathrm{E} = V_\mathrm{m}^\mathrm{E}, H_\mathrm{m}^\mathrm{E}, A_p, U_{V\mathrm{m}}^\mathrm{E}) \qquad (3)$$

by unweighted linear least-squares regressions, taking an appropriate number, $k$, of $A_i$ coefficients with guidance from F-tests of additional term [76]. The results are collected in Table 5. The root-mean-square deviation of the fit to equation (3), $s$, is defined by:

$$s = \left[\frac{1}{N-N_\mathrm{p}}\sum_{j=1}^{N}\left(F_{\mathrm{m,cal},j}^\mathrm{E} - F_{\mathrm{m,exp},j}^\mathrm{E}\right)^2\right]^{1/2} \qquad (F_\mathrm{m}^\mathrm{E} = V_\mathrm{m}^\mathrm{E}, H_\mathrm{m}^\mathrm{E}, A_p, U_{V\mathrm{m}}^\mathrm{E}) \qquad (4)$$

where $N_\mathrm{p}$ denotes the number of $A_i$ parameters in equation (3), and $j$ indexes the number $N$ of experimental data points $F_{\mathrm{m,exp},j}^\mathrm{E}$ and calculated values $F_{\mathrm{m,cal},j}^\mathrm{E}$. The value of $s$ for every fit is included in Table 5. A comparison between experimental results at 298.15 K and calculations using equation (3) is shown in Figures S1-S3.

**Table 3.** Density, $\rho$, excess molar volume, $V_m^E$, and temperature derivative of the excess molar volume, $A_p = (\partial V_m^E/\partial T)_{p,x}$, of 2-butanol (1) + glyme (2) liquid mixtures as a function of the mole fraction of 2-butanol, $x_1$, at temperature $T$ and pressure $p$ = 95 kPa [a].

| $x_1$ | $\rho$/(g cm$^{-3}$) | | | $V_m^E$/(cm$^3$ mol$^{-1}$) | | | $A_p$/(cm$^3$ mol$^{-1}$ K$^{-1}$) |
|---|---|---|---|---|---|---|---|
| | $T$/K = 293.15 | $T$/K = 298.15 | $T$/K = 303.15 | $T$/K = 293.15 | $T$/K = 298.15 | $T$/K = 303.15 | $T$/K = 298.15 |
| 2-butanol (1) + monoglyme (2) | | | | | | | |
| 0.0538 | 0.86381 | 0.85840 | 0.85297 | 0.0618 | 0.0636 | 0.0650 | 0.00032 |
| 0.1067 | 0.86040 | 0.85505 | 0.84967 | 0.1226 | 0.1259 | 0.1300 | 0.00074 |
| 0.1535 | 0.85742 | 0.85212 | 0.84679 | 0.1679 | 0.1730 | 0.1785 | 0.00106 |
| 0.2001 | 0.85442 | 0.84918 | 0.84390 | 0.2136 | 0.2192 | 0.2263 | 0.00127 |
| 0.3289 | 0.84639 | 0.84130 | 0.83618 | 0.2893 | 0.2980 | 0.3061 | 0.00168 |
| 0.4009 | 0.84189 | 0.83690 | 0.83187 | 0.3212 | 0.3297 | 0.3383 | 0.00171 |
| 0.5041 | 0.83548 | 0.83062 | 0.82572 | 0.3462 | 0.3558 | 0.3651 | 0.00189 |
| 0.6050 | 0.82939 | 0.82467 | 0.81992 | 0.3320 | 0.3410 | 0.3478 | 0.00158 |
| 0.7053 | 0.82324 | 0.81868 | 0.81408 | 0.3105 | 0.3166 | 0.3210 | 0.00105 |
| 0.8058 | 0.81744 | 0.81303 | 0.80857 | 0.2290 | 0.2330 | 0.2359 | 0.00069 |
| 0.8562 | 0.81454 | 0.81021 | 0.80583 | 0.1803 | 0.1827 | 0.1837 | 0.00034 |
| 0.9023 | 0.81187 | 0.80762 | 0.80330 | 0.1335 | 0.1338 | 0.1346 | 0.00011 |
| 0.9506 | 0.80919 | 0.80502 | 0.80077 | 0.0674 | 0.0656 | 0.0653 | −0.00021 |
| 2-butanol (1) + diglyme (2) | | | | | | | |
| 0.0627 | 0.93784 | 0.93290 | 0.92795 | 0.0604 | 0.0634 | 0.0660 | 0.00056 |
| 0.1001 | 0.93398 | 0.92905 | 0.92412 | 0.1084 | 0.1136 | 0.1168 | 0.00084 |
| 0.1578 | 0.92801 | 0.92311 | 0.91821 | 0.1550 | 0.1614 | 0.1654 | 0.00104 |
| 0.2061 | 0.92270 | 0.91782 | 0.91294 | 0.2106 | 0.2191 | 0.2248 | 0.00142 |
| 0.3018 | 0.91187 | 0.90704 | 0.90221 | 0.2833 | 0.2943 | 0.3020 | 0.00187 |
| 0.3995 | 0.90016 | 0.89539 | 0.89060 | 0.3281 | 0.3408 | 0.3523 | 0.00242 |
| 0.4998 | 0.88721 | 0.88251 | 0.87779 | 0.3634 | 0.3771 | 0.3890 | 0.00256 |
| 0.5982 | 0.87368 | 0.86907 | 0.86443 | 0.3595 | 0.3720 | 0.3831 | 0.00236 |
| 0.7005 | 0.85845 | 0.85394 | 0.84940 | 0.3368 | 0.3479 | 0.3570 | 0.00202 |
| 0.7984 | 0.84281 | 0.83841 | 0.83398 | 0.2679 | 0.2766 | 0.2824 | 0.00145 |
| 0.8475 | 0.83450 | 0.83017 | 0.82581 | 0.2209 | 0.2269 | 0.2295 | 0.00086 |
| 0.8977 | 0.82563 | 0.82138 | 0.81707 | 0.1625 | 0.1650 | 0.1673 | 0.00048 |
| 0.9483 | 0.81622 | 0.81204 | 0.80780 | 0.1007 | 0.1012 | 0.1011 | 0.00004 |
| 2-butanol (1) + triglyme (2) | | | | | | | |
| 0.1034 | 0.97449 | 0.96976 | 0.96503 | 0.0959 | 0.0990 | 0.1013 | 0.00054 |
| 0.1559 | 0.96881 | 0.96408 | 0.95937 | 0.1357 | 0.1423 | 0.1441 | 0.00084 |
| 0.1998 | 0.96379 | 0.95907 | 0.95437 | 0.1736 | 0.1814 | 0.1843 | 0.00107 |
| 0.3010 | 0.95132 | 0.94664 | 0.94196 | 0.2527 | 0.2610 | 0.2670 | 0.00143 |
| 0.4020 | 0.93767 | 0.93302 | 0.92837 | 0.2791 | 0.2897 | 0.2974 | 0.00183 |
| 0.4946 | 0.92348 | 0.91888 | 0.91426 | 0.3209 | 0.3312 | 0.3408 | 0.00199 |
| 0.6002 | 0.90518 | 0.90064 | 0.89608 | 0.3526 | 0.3632 | 0.3721 | 0.00195 |
| 0.7009 | 0.88556 | 0.88110 | 0.87660 | 0.3146 | 0.3232 | 0.3316 | 0.00170 |
| 0.7994 | 0.86337 | 0.85899 | 0.85460 | 0.2627 | 0.2706 | 0.2734 | 0.00107 |
| 0.8489 | 0.85100 | 0.84669 | 0.84233 | 0.2150 | 0.2193 | 0.2230 | 0.00080 |
| 0.9491 | 0.82275 | 0.81858 | 0.81433 | 0.0818 | 0.0809 | 0.0819 | 0.00001 |
| 2-butanol (1) + tetraglyme (2) | | | | | | | |
| 0.0537 | 1.00598 | 1.00135 | 0.99674 | 0.0386 | 0.0389 | 0.0365 | −0.00021 |
| 0.1068 | 1.00079 | 0.99616 | 0.99155 | 0.0779 | 0.0807 | 0.0804 | 0.00025 |
| 0.1559 | 0.99565 | 0.99103 | 0.98643 | 0.1183 | 0.1215 | 0.1212 | 0.00029 |
| 0.2024 | 0.99052 | 0.98591 | 0.98130 | 0.1495 | 0.1530 | 0.1567 | 0.00072 |
| 0.2994 | 0.97875 | 0.97416 | 0.96958 | 0.2143 | 0.2192 | 0.2213 | 0.00070 |
| 0.3976 | 0.96521 | 0.96065 | 0.95608 | 0.2585 | 0.2635 | 0.2684 | 0.00099 |
| 0.4979 | 0.94930 | 0.94476 | 0.94022 | 0.2759 | 0.2836 | 0.2885 | 0.00126 |
| 0.5972 | 0.93077 | 0.92628 | 0.92178 | 0.2900 | 0.2965 | 0.3007 | 0.00107 |
| 0.7004 | 0.90796 | 0.90353 | 0.89908 | 0.2672 | 0.2728 | 0.2763 | 0.00091 |
| 0.7998 | 0.88125 | 0.87690 | 0.87253 | 0.2328 | 0.2365 | 0.2369 | 0.00041 |
| 0.8474 | 0.86643 | 0.86213 | 0.85778 | 0.1989 | 0.2010 | 0.2031 | 0.00042 |
| 0.8995 | 0.84833 | 0.84409 | 0.83981 | 0.1470 | 0.1475 | 0.1460 | −0.00010 |
| 0.9492 | 0.82899 | 0.82482 | 0.82059 | 0.0732 | 0.0715 | 0.0695 | −0.00037 |

[a] Expanded uncertainties, $U$, and relative expanded uncertainties, $U_r$, with a coverage factor of 2: $U(T)$ = 0.02 K, $U(p)$ = 10 kPa, $U_r(\rho)$ = 0.002, $U(x_1)$ = 0.0010; $U(V_m^E) = 0.010\,|V_m^E|_{max}$ + 0.005 cm$^3$ mol$^{-1}$, $U(A_p) = 0.010\,|A_p|_{max}$ + 0.0002 cm$^3$ mol$^{-1}$ K$^{-1}$.

**Table 4.** Excess molar enthalpy, $H_{\mathrm{m}}^{\mathrm{E}}$, of 2-butanol (1) + glyme (2) liquid mixtures as function of the mole fraction of 2-butanol, $x_1$, at temperature $T$ = 298.15 K and pressure $p$ = 0.95 kPa [a].

| $x_1$ | $H_{\mathrm{m}}^{\mathrm{E}}/(\mathrm{J\ mol^{-1}})$ | $x_1$ | $H_{\mathrm{m}}^{\mathrm{E}}/(\mathrm{J\ mol^{-1}})$ | $x_1$ | $H_{\mathrm{m}}^{\mathrm{E}}/(\mathrm{J\ mol^{-1}})$ | $x_1$ | $H_{\mathrm{m}}^{\mathrm{E}}/(\mathrm{J\ mol^{-1}})$ |
|---|---|---|---|---|---|---|---|
| 2-butanol (1) + monoglyme (2) | | 2-butanol (1) + diglyme (2) | | 2-butanol (1) + triglyme (2) | | 2-butanol (1) + tetraglyme (2) | |
| 0.0542 | 312 | 0.0536 | 299 | 0.0597 | 364 | 0.0761 | 463 |
| 0.1030 | 547 | 0.1002 | 575 | 0.1028 | 635 | 0.1075 | 662 |
| 0.1517 | 762 | 0.1485 | 824 | 0.1537 | 912 | 0.1553 | 952 |
| 0.2019 | 968 | 0.2028 | 1079 | 0.1992 | 1142 | 0.2065 | 1226 |
| 0.3069 | 1260 | 0.3006 | 1443 | 0.3053 | 1604 | 0.3059 | 1682 |
| 0.3986 | 1427 | 0.3990 | 1698 | 0.4007 | 1907 | 0.4023 | 2042 |
| 0.5014 | 1504 | 0.4950 | 1846 | 0.5027 | 2096 | 0.4977 | 2253 |
| 0.6014 | 1461 | 0.6067 | 1843 | 0.6000 | 2137 | 0.5912 | 2349 |
| 0.6995 | 1306 | 0.6975 | 1708 | 0.6873 | 2031 | 0.7037 | 2218 |
| 0.8036 | 1001 | 0.7983 | 1371 | 0.8041 | 1606 | 0.8044 | 1851 |
| 0.8448 | 837 | 0.8489 | 1123 | 0.8464 | 1389 | 0.8502 | 1570 |
| 0.8922 | 607 | 0.8905 | 878 | 0.9075 | 959 | 0.8928 | 1249 |
| 0.9367 | 370 | 0.9444 | 482 | 0.9461 | 599 | 0.9451 | 725 |

[a] Expanded uncertainties, $U$, and relative expanded uncertainties, $U_{\mathrm{r}}$, with a coverage factor of 2: $U(T)$ = 0.02 K, $U(p)$ = 10 kPa, $U(x_1)$ = 0.0010, $U_{\mathrm{r}}(H_{\mathrm{m}}^{\mathrm{E}})$ = 0.030.

**Table 5**. Parameters $A_i$ of equation (3) and root-mean-square deviation of the fit, $s$, equation (4), for the excess molar volume, $V_{\mathrm{m}}^{\mathrm{E}}$, the temperature derivative of the excess molar volume, $A_p = (\partial V_{\mathrm{m}}^{\mathrm{E}}/\partial T)_{p,x}$, and the excess molar enthalpy, $H_{\mathrm{m}}^{\mathrm{E}}$, of 2-butanol (1) + glyme (2) liquid mixtures at temperature $T$ and pressure $p$ = 95 kPa.

| Property | $T$/K | $A_0$ | $A_1$ | $A_2$ | $A_3$ | $s$ |
|---|---|---|---|---|---|---|
| | | 2-butanol (1) + monoglyme (2) | | | | |
| $V_{\mathrm{m}}^{\mathrm{E}}/(\mathrm{cm^3\ mol^{-1}})$ | 293.15 | 1.381 | 0.15 | | | 0.005 |
| | 298.15 | 1.415 | 0.13 | | | 0.005 |
| | 303.15 | 1.446 | 0.11 | | | 0.005 |
| $A_p/(\mathrm{cm^3\ mol^{-1}\ K^{-1}})$ | 298.15 | 0.0073 | −0.0034 | −0.004 | | 0.00011 |
| $H_{\mathrm{m}}^{\mathrm{E}}/(\mathrm{J\ mol^{-1}})$ | 298.15 | 6023 | 294 | 260 | | 8 |
| $U_{V\mathrm{m}}^{\mathrm{E}}/(\mathrm{J\ mol^{-1}})$ | 298.15 | 5556 | 266 | 261 | | 0.3 |
| | | 2-butanol (1) + diglyme (2) | | | | |
| $V_{\mathrm{m}}^{\mathrm{E}}/(\mathrm{cm^3\ mol^{-1}})$ | 293.15 | 1.46 | 0.36 | | | 0.006 |
| | 298.15 | 1.51 | 0.36 | | | 0.006 |
| | 303.15 | 1.55 | 0.35 | | | 0.005 |
| $A_p/(\mathrm{cm^3\ mol^{-1}\ K^{-1}})$ | 298.15 | 0.0102 | −0.0007 | −0.005 | | 0.00014 |
| $H_{\mathrm{m}}^{\mathrm{E}}/(\mathrm{J\ mol^{-1}})$ | 298.15 | 7394 | 1597 | 477 | | 9 |
| $U_{V\mathrm{m}}^{\mathrm{E}}/(\mathrm{J\ mol^{-1}})$ | 298.15 | 6857 | 59 | 456 | | 4.5 |
| | | 2-butanol (1) + triglyme (2) | | | | |
| $V_{\mathrm{m}}^{\mathrm{E}}/(\mathrm{cm^3\ mol^{-1}})$ | 293.15 | 1.33 | 0.46 | | | 0.01 |
| | 298.15 | 1.37 | 0.45 | | | 0.01 |
| | 303.15 | 1.4 | 0.46 | | | 0.0090 |
| $A_p/(\mathrm{cm^3\ mol^{-1}\ K^{-1}})$ | 298.15 | 0.0081 | 0.0003 | −0.004 | | 0.00011 |
| $H_{\mathrm{m}}^{\mathrm{E}}/(\mathrm{J\ mol^{-1}})$ | 298.15 | 8375 | 2333 | 971 | 679 | 8 |
| $U_{V\mathrm{m}}^{\mathrm{E}}/(\mathrm{J\ mol^{-1}})$ | 298.15 | 7867 | 2229 | 998 | 675 | 0.3 |
| | | 2-butanol (1) + tetraglyme (2) | | | | |
| $V_{\mathrm{m}}^{\mathrm{E}}/(\mathrm{cm^3\ mol^{-1}})$ | 293.15 | 1.15 | 0.42 | | | 0.008 |
| | 298.15 | 1.18 | 0.42 | | | 0.008 |
| | 303.15 | 1.19 | 0.42 | | | 0.007 |
| $A_p/(\mathrm{cm^3\ mol^{-1}\ K^{-1}})$ | 298.15 | 0.0049 | −0.0004 | −0.006 | | 0.00019 |
| $H_{\mathrm{m}}^{\mathrm{E}}/(\mathrm{J\ mol^{-1}})$ | 298.15 | 9046 | 3158 | 1473 | 1180 | 8 |
| $U_{V\mathrm{m}}^{\mathrm{E}}/(\mathrm{J\ mol^{-1}})$ | 298.15 | 8592 | 3058 | 1507 | 1190 | 0.3 |

# 4. Models

## 4.1. DISQUAC

DISQUAC is based on Guggenheim's rigid lattice theory [77]. We briefly summarize the main features of the model. (i) The geometrical parameters (total molecular volumes, $r_i$, surfaces, $q_i$, and molecular surface fractions, $\alpha_i$) of the mixture components are calculated additively based on the group volumes $R_G$ and surfaces $Q_G$ recommended by Bondi [78]. The volume $R_{\mathrm{CH4}}$ and surface $Q_{\mathrm{CH4}}$ of methane are taken arbitrarily as volume and surface units [79]. The geometrical parameters for the groups involved in this research are available in the literature [79–81]. (ii) Two interactional contributions exist to the excess molar functions: a dispersive (DIS) term, related to London dispersion forces, and a quasichemical (QUAC) term due to the anisotropy of the field forces created by the solution molecules. For the excess molar Gibbs energy, $G_{\mathrm{m}}^{\mathrm{E}}$, a combinatorial term, $G_{\mathrm{m}}^{\mathrm{E,COMB}}$, represented by the Flory-Huggins equation [79,82] must be included:

$$G_{\mathrm{m}}^{\mathrm{E}} = G_{\mathrm{m}}^{\mathrm{E,COMB}} + G_{\mathrm{m}}^{\mathrm{E,DIS}} + G_{\mathrm{m}}^{\mathrm{E,QUAC}} \quad (5)$$

$$H_{\mathrm{m}}^{\mathrm{E}} = H_{\mathrm{m}}^{\mathrm{E,DIS}} + H_{\mathrm{m}}^{\mathrm{E,QUAC}} \quad (6)$$

(iii) The interaction parameters depend on the molecular structure. (iv) The coordination number $z$ = 4 is used for all the polar contacts. This shortcoming of DISQUAC is partially removed using interaction parameters dependent on the molecular structure. (v) No volume change exists upon mixing ($V_{\mathrm{m}}^{\mathrm{E}}$ = 0). The equations required to calculate the DIS and QUAC contributions to $G_{\mathrm{m}}^{\mathrm{E}}$ and $H_{\mathrm{m}}^{\mathrm{E}}$ can be found elsewhere [82,83]. The temperature dependence of the interaction parameters is expressed in terms of the DIS and QUAC interchange coefficients [82,83], $C_{st,l}^{\mathrm{DIS}}, C_{st,l}^{\mathrm{QUAC}}$ where $s \neq t$ are two contact surfaces present in the mixture and $l$ = 1 (Gibbs energy; $C_{st,1}^{\mathrm{DIS,QUAC}} = g_{st}^{\mathrm{DIS,QUAC}}(T_0)/RT_0$); $l$ = 2 (enthalpy, $C_{st,2}^{\mathrm{DIS,QUAC}} = h_{st}^{\mathrm{DIS,QUAC}}(T_0)/RT_0$)); $l$ = 3 (heat capacity, $C_{st,3}^{\mathrm{DIS,QUAC}} = c_{pst}^{\mathrm{DIS,QUAC}}(T_0)/R$)). $T_0$ = 298.15 K is the scaling temperature and $R$ is the gas constant.

## 4.2. UNIFAC (Dortmund)

This version of UNIFAC [29] differs from original UNIFAC [84] by the combinatorial term and the temperature dependence of the interaction parameters. The equations to calculate $G_{\mathrm{m}}^{\mathrm{E}}$ and $H_{\mathrm{m}}^{\mathrm{E}}$ [83] are derived from the fundamental equation for the activity coefficient $\gamma_i$ of component $i$:

$$\ln \gamma_i = \ln \gamma_i^{\mathrm{COMB}} + \ln \gamma_i^{\mathrm{RES}} \quad (7)$$

where $\ln \gamma_i^{\mathrm{COMB}}$ and $\ln \gamma_i^{\mathrm{RES}}$ are, respectively, the combinatorial and residual terms. In UNIFAC (Dortmund), the mixtures studied are built by three main groups: No 1, ($CH_3$); No 5, OH, for primary, secondary and tertiary alcohols, and No 15, ether. The main groups include different subgroups which have different geometrical parameters, but the subgroups within the same main group have identical group energy-interaction parameters.

The geometrical parameters, the relative van der Waals volumes, and the relative van der Waals surfaces of the different subgroups are not calculated from molecular parameters like in the original UNIFAC but fit together with the interaction parameters to the experimental values of the thermodynamic properties considered. In this study, the geometrical and interaction parameters for the involved groups were taken from the literature and used without modification [30].

## 4.3. ERAS

The Extended Real Associated Solution model (ERAS) [24] combines the Real Association Solution Model [83,85] with Flory's equation of state [25]. Important features of the model are the following. (i) The excess molar functions ($F_{\mathrm{m}}^{\mathrm{E}} = H_{\mathrm{m}}^{\mathrm{E}}$, $V_{\mathrm{m}}^{\mathrm{E}}$) are the result of two contributions. The chemical one, $F_{\mathrm{m,chem}}^{\mathrm{E}}$, is linked to hydrogen bonding; the physical contribution, $F_{\mathrm{m,phys}}^{\mathrm{E}}$, arises from nonpolar Van der Waals interactions and free volume effects. Expressions for $H_{\mathrm{m}}^{\mathrm{E}}$ and $V_{\mathrm{m}}^{\mathrm{E}}$ can be found elsewhere [83] (ii) It is assumed that only consecutive linear associations occur. Thus, self-association of a compound is described by a chemical equilibrium constant ($K_i$) independent of the chain length of the self-associated species. In the present study, only 2-alkanols (molecules A) are self-associated, which is described by the equation:

$$\mathrm{A}_m + \mathrm{A} \overset{K_{\mathrm{A}}}{\leftrightarrow} \mathrm{A}_{m+1} \tag{8}$$

with $m$ ranging from 1 to ∞. The cross-association between a self-associated component $\mathrm{A}_m$ and a non-associated compound B (ether) is described by:

$$\mathrm{A}_m + \mathrm{B} \overset{K_{\mathrm{AB}}}{\leftrightarrow} \mathrm{A}_m\mathrm{B} \tag{9}$$

The cross-association constants, $K_{\mathrm{AB}}$, are also assumed independent of $m$. Reactions described by equations (8) and (9) are characterized, respectively, by the molar enthalpies of intermolecular hydrogen-bonding $\Delta h_{\mathrm{A}}^*$, and $\Delta h_{\mathrm{AB}}^*$, and by negative molar hydrogen-bonding volumes, $\Delta v_{\mathrm{A}}^*$, and $\Delta v_{\mathrm{AB}}^*$. The latter are needed to consider the decrease of the core volume of the molecules upon the formation of multimers. The equilibrium constants depend on temperature according to the $\Delta h_i^*$ values and the Van't Hoff equation. (iii) The $F_{\mathrm{m,phys}}^{\mathrm{E}}$ term is obtained from the Flory's equation of state and depends on a physical interaction parameter, $X_{\mathrm{AB}}$ [25]. It is assumed that the mentioned equation of state is valid for pure compounds and for mixtures [86,87]:

$$\frac{\bar{p}_i \bar{V}_i}{\bar{T}_i} = \frac{\bar{V}_i^{1/3}}{\bar{V}_i^{1/3} - 1} - \frac{1}{\bar{V}_i \bar{T}_i} \tag{10}$$

where $i$ = A, B or M (mixture). In equation (10), $\bar{V}_i = V_{\mathrm{m},i}/V_{\mathrm{m},i}^*$; $\bar{p}_i = p/p_i^*$; $\bar{T}_i = T/T_i^*$ are the reduced properties for volume, pressure and temperature, respectively. The pure component reduction parameters ($V_{\mathrm{m},i}^*, p_i^*, T_i^*$) are obtained from $p$-$V$-$T$ data ($\rho$, $\alpha_{pi}$ and $\kappa_{Ti}$) and association parameters [86,87]. For mixtures, the reduction parameters $p_{\mathrm{M}}^*$ and $T_{\mathrm{M}}^*$ are calculated from the application of certain mixing rules [86,87]. The total relative molecular volumes and surfaces of the compounds were calculated additively according to Bondi's method [78].

# 5. Estimation of interaction parameters

## 5.1. DISQUAC

In the framework of DISQUAC, 2-alkanol + linear ether mixtures are built by three surfaces: (i) type a, aliphatic ($CH_3$, $CH_2$, or CH in 2-alkanol or ethers); (ii) type e, O in oxaalkanes; (iii) type h, OH in 2-alkanols. The three surfaces generate three contacts: (a,e), (a,h) and (e,h). The (a,e) and the (a,h) contacts are described by DIS and QUAC interaction parameters and are known from the study of ether [80] or 2-alkanol [88] + alkane systems. In consequence, only the interaction parameters for the (e,h) contacts must be determined. The general procedure applied when estimating the DISQUAC interaction parameters have been given elsewhere [82,83]. Due to the lack of the required experimental data, we have used $C_{\mathrm{eh},3}^{\mathrm{DIS/QUAC}} = 0$ for all the systems. The fitted interchange coefficients are listed in Table 6.

**Table 6**. Dispersive (DIS) and quasichemical (QUAC) interchange coefficients, $C_{\mathrm{eh},l}^{\mathrm{DIS}}$ and $C_{\mathrm{eh},l}^{\mathrm{QUAC}}$, for (e,h) contacts[a] in (2-alkanol + linear ether) mixtures ($l$ = 1, Gibbs energy; $l$ = 2, enthalpy); $u$ denotes the number of O atoms in the glyme, for DPE, $u$ = 1.

| 2-alkanol[b] | ether | $C_{\mathrm{eh},1}^{\mathrm{DIS}}$ | $C_{\mathrm{eh},2}^{\mathrm{DIS}}$ | $C_{\mathrm{eh},1}^{\mathrm{QUAC}}$ | $C_{\mathrm{eh},2}^{\mathrm{QUAC}}$ |
|---|---|---|---|---|---|
| 2PrOH | $u$ = 1 | 7.2 | 7.2 | 1 | 8.9 |
| | $u$ = 2 | 7[c] | 27.6 | 1 | 0.9 |
| | $u$ = 3 | 7[c] | 27.6 | 1 | 0.15 |
| | $u$ = 4 | 7[c] | 27.6 | 1 | 0.15 |
| | $u$ = 5 | 7[c] | 27.6 | 1 | 0.15 |
| 2BuOH | $u$ = 1 | 4.5 | 8.8 | 1 | 8.9 |
| | $u$ = 2 | 7[c] | 31 | 1 | 0.9 |
| | $u$ = 3 | 7[c] | 31 | 1 | 0.15 |
| | $u$ = 4 | 7[c] | 31 | 1 | 0.15 |
| | $u$ = 5 | 7[c] | 31 | 1 | 0.15 |

[a] Type a, $CH_3$, $CH_2$, or CH in linear ethers or 2-alkanols; type h, OH in 2-alkanols.
[b] Symbols: 2PrOH, 2-propanol; 2BuOH, 2-butanol.
[c] Estimated values from the study of similar systems [6,7,18,20,89–91].

## 5.2. ERAS

The values of the ERAS parameters $K_\mathrm{A}$, $\Delta h_\mathrm{A}^*$, $\Delta v_\mathrm{A}^*$ for 2-alkanols, and of their reduction parameters are known from the study of 2-alkanol + *n*-alkane mixtures [89,92]. For the pure compounds considered in this work, the values of the mentioned parameters are listed in Table S1. 2-Alkanol + DPE or + glyme systems are characterized by the binary parameters $K_\mathrm{AB}$, $\Delta h_\mathrm{AB}^*$, $\Delta v_\mathrm{AB}^*$ and $X_\mathrm{AB}$ (Table 7), which were fitted using data on $H_\mathrm{m}^\mathrm{E}$ and $V_\mathrm{m}^\mathrm{E}$ available in the literature (this work, [28,89,90]).

**Table 7**. ERAS parameters[a] for 2-alkanol + linear ether mixtures at temperature $T$ = 298.15 K and atmospheric pressure.

| System[b] | $K_{AB}$ | $\Delta h^{*}_{AB}$/kJ mol$^{-1}$ | $\Delta v^{*}_{AB}$/cm$^3$ mol$^{-1}$ | $X_{AB}$/J cm$^{-3}$ |
|---|---|---|---|---|
| 2PrOH + DPE | 7.5 | −13.3 | −6.8 | 6.7 |
| 2PrOH + monoglyme | 18.2 | −13.3 | −6.5 | 12.5 |
| 2PrOH + diglyme | 22.6 | −13.3 | −6.5 | 16.8 |
| 2PrOH + triglyme | 31.3 | −13.3 | −6.1 | 17.4 |
| 2PrOH + tetraglyme | 42.2 | −13.3 | −5.9 | 17.2 |
| 2BuOH + DPE | 2.7 | −13.3 | −9.3 | 12.5 |
| 2BuOH + monoglyme | 10.7 | −13.3 | −6.4 | 23.6 |
| 2BuOH + diglyme | 13.3 | −13.3 | −6.4 | 28.4 |
| 2BuOH + triglyme | 18.1 | −13.3 | −6.1 | 28.7 |
| 2BuOH + tetraglyme | 21.3 | −13.3 | −5.9 | 27.8 |

[a] $K_{AB}$, equilibrium constant; $\Delta h^{*}_{AB}$, molar enthalpy of intermolecular hydrogen-bonding; $\Delta v^{*}_{AB}$, molar hydrogen-bonding volume; $X_{AB}$, physical parameter.
[b] Symbols: 2PrOH, 2-propanol; 2BuOH, 2-butanol, DPE, dipropyl ether.

# 6. Results from models

Results on $H^{E}_{m}$ from the DISQUAC, UNIFAC (Dortmund) and ERAS models are collected in Table 8 (see Figures 1,2), which also contains relative deviations for $H^{E}_{m}$:

$$\mathrm{dev}\left(H^{E}_{m}\right) = \left[\frac{1}{N}\sum\left(\frac{H^{E}_{m,exp} - H^{E}_{m,calc}}{H^{E}_{m,exp}(x_1 = 0.5)}\right)^2\right]^{1/2} \quad (11)$$

We note that, for glyme systems, DISQUAC improves largely the $H^{E}_{m}$ results provided by UNIFAC (Dortmund) or ERAS. A direct graphical comparison between results obtained from the applied models is shown in Figures 3 and S4. The latter shows that DISQUAC and ERAS provide similar results for mixtures with DPE (Table 8). DISQUAC results on VLE are shown in Table 9 (see Figure 4), which lists the relative standard deviations for pressure defined by:

$$\sigma_r(P) = \left[\frac{1}{N}\sum\left(\frac{P_{exp} - P_{calc}}{P_{exp}}\right)^2\right]^{1/2} \quad (12)$$

Results on $V^{E}_{m}$ provided by ERAS are collected in Table 10 and Figure 5. The model yields fairly good results for this quantity.

**Table 8**. Excess molar enthalpies, $H_m^E$, at equimolar composition, temperature $T$ = 298.15 K and atmospheric pressure for 2-alkanol (1) + linear ether (2) mixtures. Comparison of experimental results (exp) with Redlich-Kister polynomials (equation (3)) with coefficients from Table 5 (RK), DISQUAC (DQ), UNIFAC (Dortmund) (UNIF) and ERAS results.

| 2-alkanol[a] | $u$[b] | $N$[c] | $H_m^E$/J mol$^{-1}$ | | dev($H_m^E$)[d] | | | | Ref. |
|---|---|---|---|---|---|---|---|---|---|
| | | | Exp. | DQ[c] | RK | DQ[c] | UNIF | ERAS | |
| 2PrOH | 1 | 14 | 958 | 972 | 0.012 | 0.018 | 0.281 | 0.013 | [90] |
| | 2 | 13 | 1323 | 1321 | 0.002 | 0.017 | 0.193 | 0.156 | [28] |
| | 3 | 13 | 1624 | 1624 | 0.008 | 0.038 | 0.196 | 0.156 | [28] |
| | 4 | 13 | 1846 | 1790 | 0.008 | 0.034 | 0.169 | 0.161 | [28] |
| | 5 | 13 | 1995 | 1987 | 0.006 | 0.036 | 0.175 | 0.168 | [28] |
| 2BuOH | 1 | 15 | 1056 | 1071 | 0.021 | 0.018 | 0.361 | 0.037 | [89] |
| | 2 | 13 | 1506 | 1504 | 0.006 | 0.015 | 0.172 | 0.155 | this work |
| | 3 | 13 | 1848 | 1809 | 0.005 | 0.021 | 0.200 | 0.156 | this work |
| | 4 | 13 | 2094 | 2093 | 0.004 | 0.005 | 0.209 | 0.154 | this work |
| | 5 | 13 | 2261 | 2322 | 0.004 | 0.021 | 0.212 | 0.164 | this work |

[a] Symbols: 2PrOH, 2-propanol; 2BuOH, 2-butanol.
[b] Number of O atoms in the glyme, for DPE, $u$ = 1.
[c] Number of data points.
[d] Equation (11).

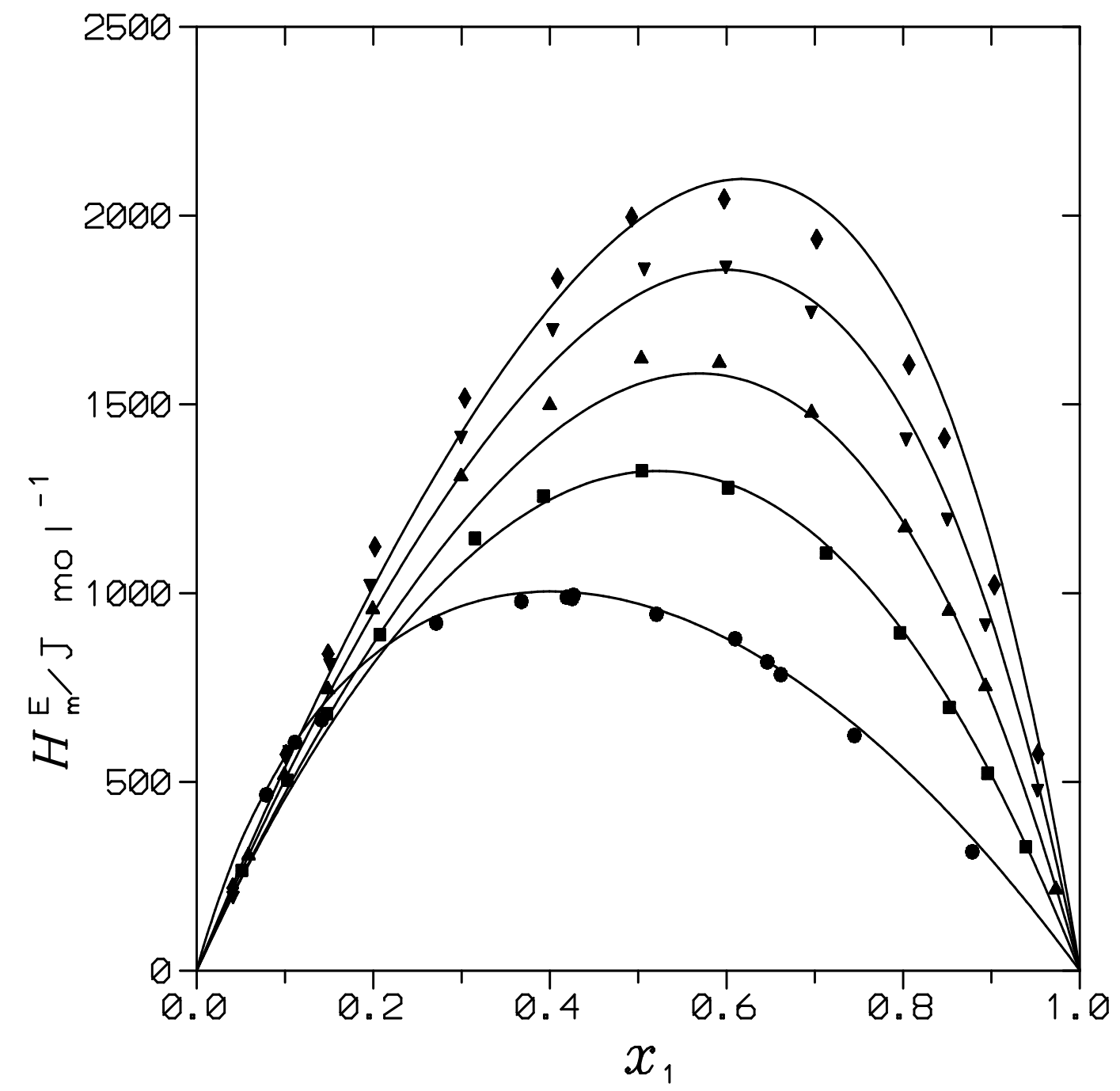

**Figure 1**. $H_m^E$ for 2-propanol (1) + linear ether (2) mixtures at temperature $T$ = 298.15 K and atmospheric pressure. Points, experimental results (this work, [90]): (●), dipropyl ether; (■), 2,5-dioxahexane; (▲), 2,5,8-trioxanonane; (▼), 2,5,8,11-tetraoxadodecane; (◆), 2,5,8,11,14-pentaoxapentadecane. Solid lines, DISQUAC calculations with interchange coefficients listed in Table 6.

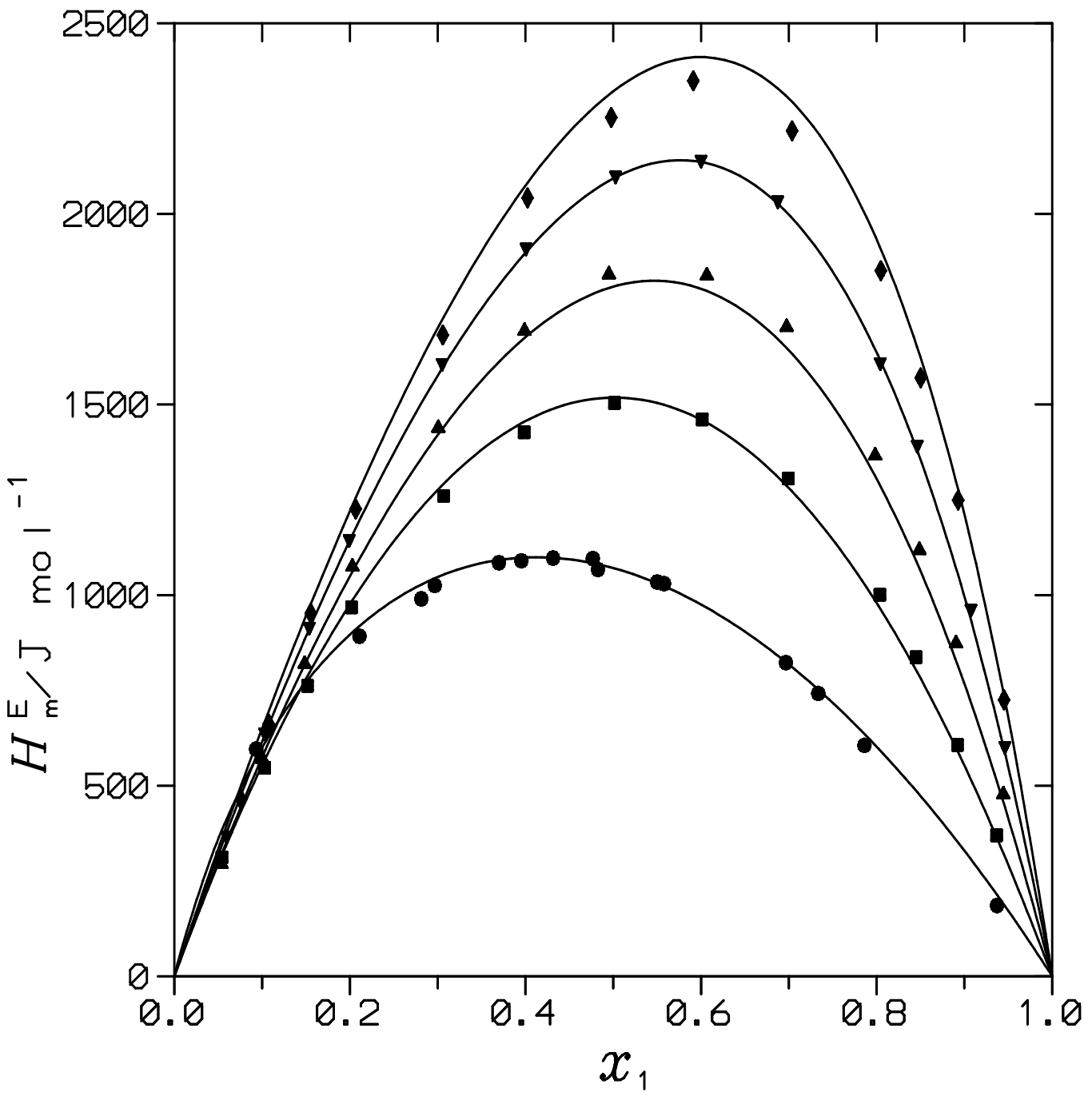

**Figure 2**. $H_m^E$ for 2-butanol (1) + linear ether (2) mixtures at temperature $T$ = 298.15 K and atmospheric pressure. Points, experimental results (this work, [89]): (●), dipropyl ether; (■), 2,5-dioxahexane; (▲), 2,5,8-trioxanonane; (▼), 2,5,8,11-tetraoxadodecane; (◆), 2,5,8,11,14-pentaoxapentadecane. Solid lines, DISQUAC calculations with interchange coefficients listed in Table 6.

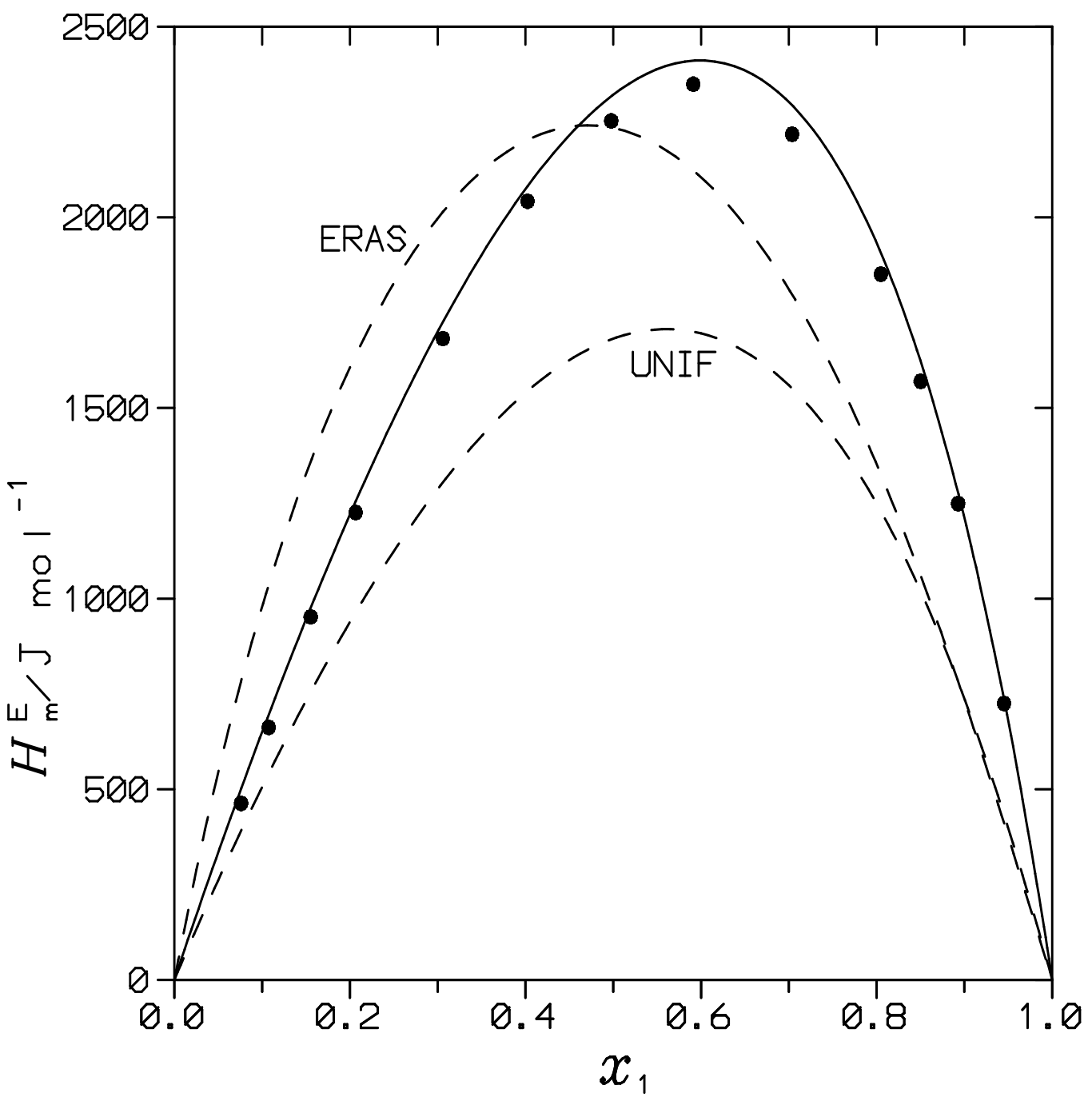

**Figure 3**. $H_m^E$ for the 2-butanol (1) + 2,5,8,11,14-pentaoxpentadecane (2) mixture at temperature $T$ = 298.15 K and atmospheric pressure. Points, experimental results (this work). Solid lines, DISQUAC calculations with interchange coefficients listed in Table 6. Dashed lines, results from the application of the ERAS and UNIFAC (Dortmund) using, respectively, interaction parameters collected in Table 7 or from the literature [30].

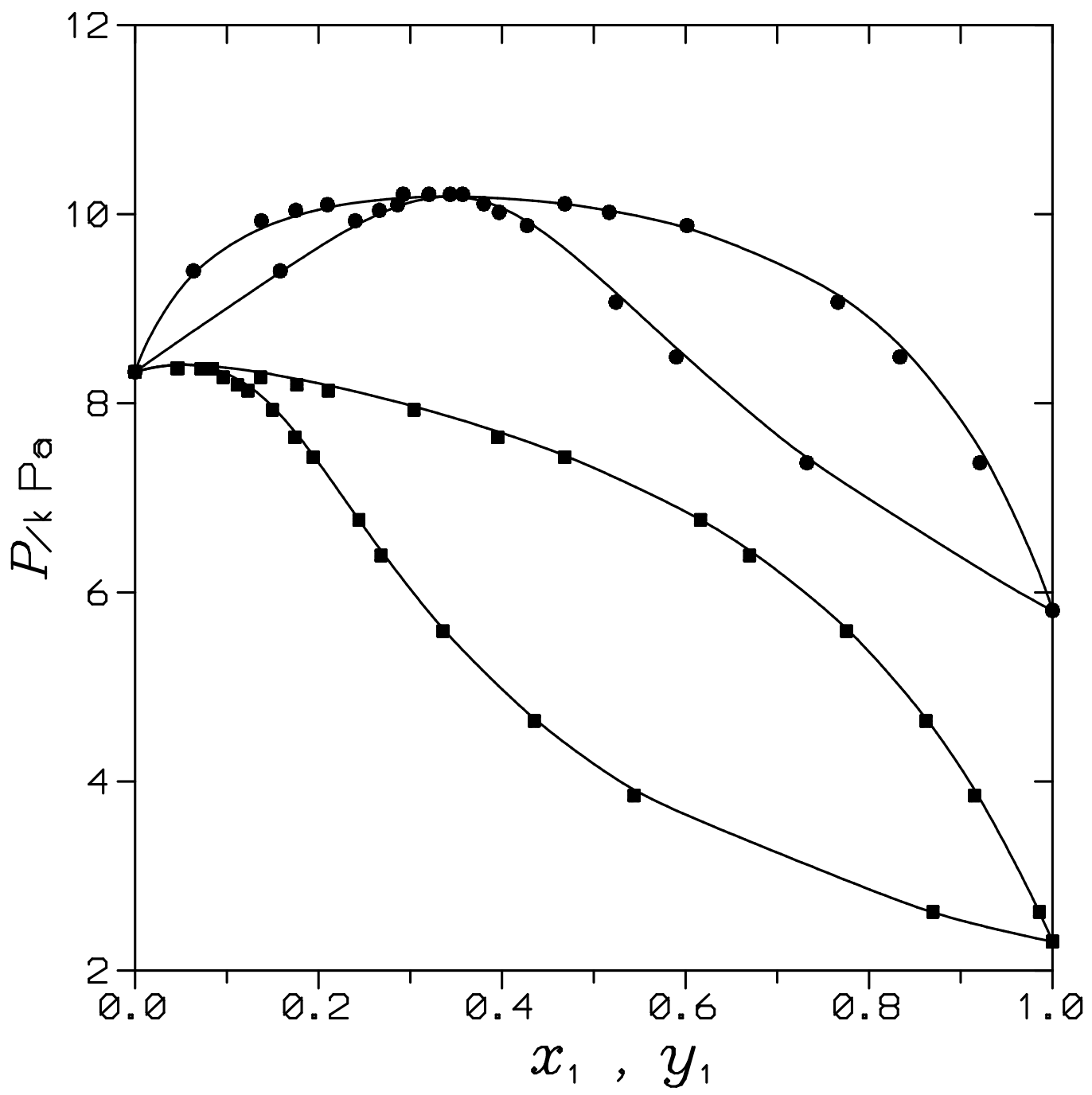

**Figure 4**. VLE for the systems 2-propanol (1), or 2-butanol (1) + dipropyl ether (2) at temperature $T$ = 298.15 K. Points, experimental results: (●), 2-propanol [90]; (■), 2-butanol [89]. Solid lines, DISQUAC calculations with interchange coefficients listed in Table 6.

**Table 9**. Excess molar Gibbs energies, $G_{\mathrm{m}}^{\mathrm{E}}$, at equimolar composition and temperature $T$ for 2-alkanol (1) + dipropyl ether (2) mixtures. Coordinates of the azeotropes ($x_{1\mathrm{az}}$, $P_{\mathrm{az}}$) are also given.

| 2-alkanol | $T$/K | $N^a$ | $G_{\mathrm{m}}^{\mathrm{E}}$/J mol$^{-1}$ | | $\sigma_{\mathrm{r}}(P)^{b}$ | | $x_{1\mathrm{az}}$ | | $P_{\mathrm{az}}$/kPa | | Ref. |
|---|---|---|---|---|---|---|---|---|---|---|---|
| | | | Exp. | DQ[c] | Exp. | DQ[c] | Exp. | DQ[c] | Exp. | DQ[c] | |
| 2-propanol | 278.15 | 12 | 854 | 861 | 0.005 | 0.012 | 0.255 | 0.262 | 3.31 | 3.32 | [90] |
| | 298.15 | 12 | 849 | 860 | 0.006 | 0.007 | 0.337 | 0.342 | 10.21 | 10.19 | [90] |
| | 323.15 | 12 | 837 | 838 | 0.002 | 0.006 | 0.446 | 0.460 | 33.49 | 33.50 | [90] |
| 2-butanol | 278.15 | 14 | 767 | 749 | 0.004 | 0.014 | | | | | [89] |
| | 298.15 | 14 | 725 | 733 | 0.004 | 0.007 | 0.037 | 0.050 | 8.45 | 8.38 | [89] |
| | 323.15 | 14 | 684 | 693 | 0.002 | 0.005 | 0.117 | 0.120 | 25.82 | 25.77 | [89] |

[a] Number of data points.

[b] Equation (12); DISQUAC values obtained with interaction parameters from Table 6.

**Table 10**. Excess molar volumes, $V_m^E$, for 2-alkanol (1) + linear ether (2) mixtures at equimolar composition, temperature $T$ = 298.15 K and atmospheric pressure. Comparison of experimental values (Exp.) with ERAS results using interaction parameters listed in Table 7.

| System[a] | $V_m^E/(cm^3\ mol^{-1})$ | | Ref. |
|---|---|---|---|
| | Exp. | ERAS | |
| 2PrOH + dipropyl ether | −0.027 | −0.037 | [90] |
| 2PrOH + monoglyme | 0.176 | 0.187 | [28] |
| 2PrOH + diglyme | 0.105 | 0.098 | [28] |
| 2PrOH + triglyme | 0.023 | 0.023 | [28] |
| 2PrOH + tetraglyme | −0.044 | −0.056 | [28] |
| 2BuOH + dipropyl ether | −0.073 | −0.097 | [89] |
| 2BuOH + monoglyme | 0.354 | 0.364 | this work |
| 2BuOH + diglyme | 0.377 | 0.382 | this work |
| 2BuOH + triglyme | 0.342 | 0.321 | this work |
| 2BuOH + tetraglyme | 0.295 | 0.277 | this work |

[a] Symbols: 2PrOH, 2-propanol; 2BuOH, 2-butanol.

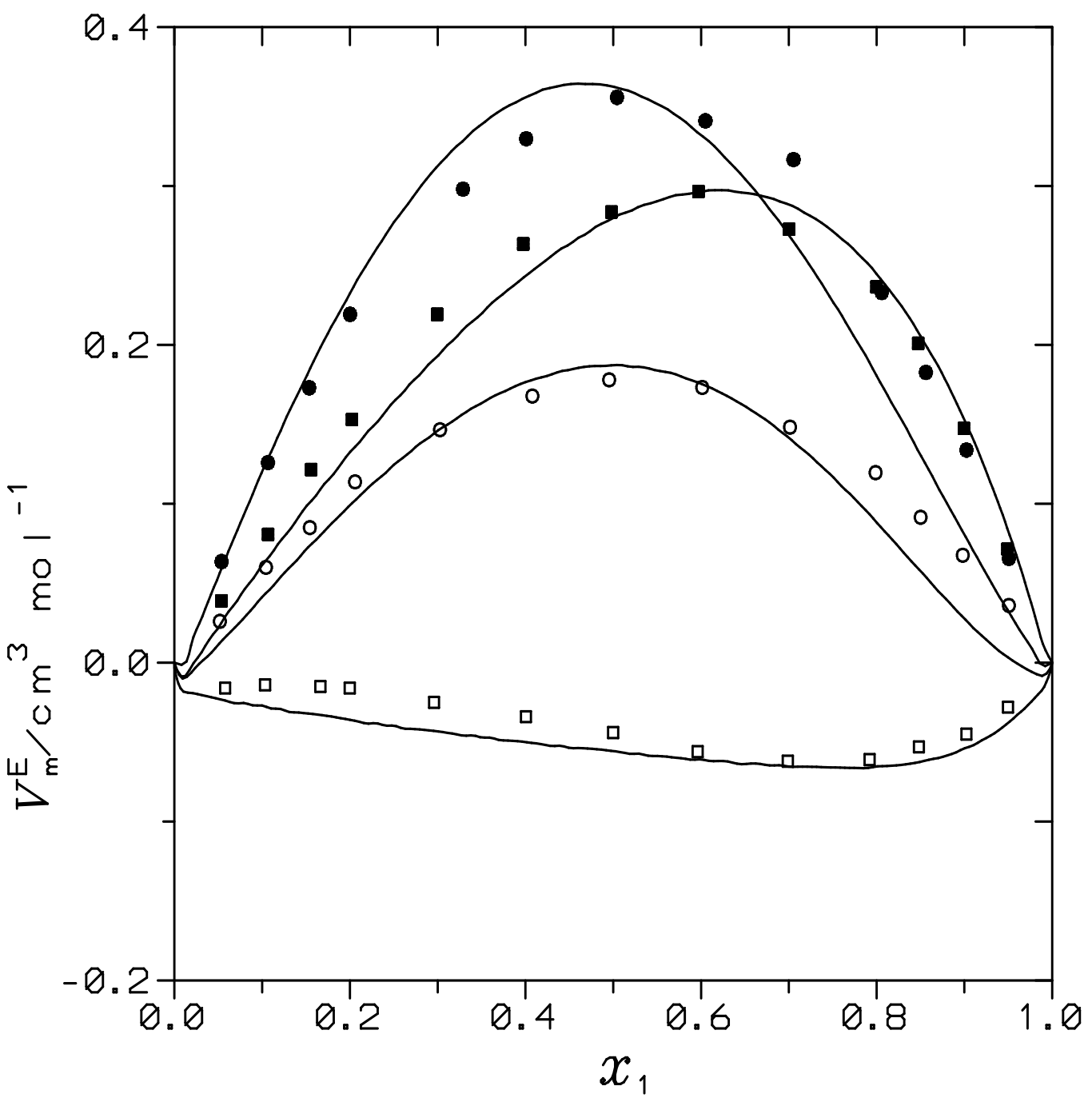

**Figure 5**. $V_m^E$ for 2-alkanol (1) + linear polyether (2) mixtures at temperature $T$ = 298.15 K and atmospheric pressure. Comparison of experimental results (points) with ERAS calculations using interaction parameters from Table 7 (solid lines): (●), 2-butanol (1) + 2,5-dioxahexane (2); (■), 2-butanol (1) + 2,5,8,11,14-pentaoxapentadecane (2) (this work); (○), 2-propanol (1) + 2,5-dioxahexane (2); (□), 2-propanol (1) + 2,5,8,11,14-pentaoxapentadecane (2) [28].

# 7. Discussion

Below, the values of the excess properties are given at equimolar composition and temperature $T$ = 298.15 K.

## 7.1. Enthalpic data

Firstly, we note that the systems 2PrOH or 2BuOH + DPE show positive azeotropes [89,90] (Figure 4) and, therefore, are characterized by positive deviations from the Raoult's law. On the other hand, $H_{\mathrm{m}}^{\mathrm{E}}$ values of 2-alkanol + linear ether mixtures are positive and higher than those of the 2-alkanol + heptane systems. For example, $H_{\mathrm{m}}^{\mathrm{E}}$(2BuOH)/J mol$^{-1}$ = 906 (heptane) [93]; 1056 (DPE) [89]; 1506 (2,5-dioxahexane). This means that ethers are good breakers of the alcohol network, and that interactions between like molecules are dominant. In addition, the $H_{\mathrm{m}}^{\mathrm{E}}$ values of 2-alkanol + glyme mixtures are very large and not far from those of glyme + heptane systems (Figure 6). In the case of solutions with 2BuOH, the values are even higher. This reveals that the contribution to $H_{\mathrm{m}}^{\mathrm{E}}$ from the breaking of glyme-glyme interactions is also determinant and that 2-alkanol + glyme mixtures are mainly characterized by physical interactions. It is remarkable that the $H_{\mathrm{m}}^{\mathrm{E}}$ curves become skewed towards higher $x_1$ values when, in systems with a given 2-alkanol, $u$ increases (Figures 1 and 2), which suggests that effects related to 2-alkanol self-association are of minor importance in such solutions. To gain a deeper understanding on this matter, we consider now the situation of the 2-alkanol + DPE systems within the $G_{\mathrm{m}}^{\mathrm{E}}$ *vs* $H_{\mathrm{m}}^{\mathrm{E}}$ diagram [94,95]. We have, $G_{\mathrm{m}}^{\mathrm{E}}/H_{\mathrm{m}}^{\mathrm{E}}$(DPE) = 849/958 [90] = 0.89 (2PrOH); 725/1056 [89] = 0.69 (2BuOH). It is well known that systems where dispersive interactions are preponderant are usually situated between the lines $G_{\mathrm{m}}^{\mathrm{E}}$ = (1/3)$H_{\mathrm{m}}^{\mathrm{E}}$ and $G_{\mathrm{m}}^{\mathrm{E}}$ = (1/2)$H_{\mathrm{m}}^{\mathrm{E}}$ of the diagram, while those characterized by dipolar interactions are placed between the lines $G_{\mathrm{m}}^{\mathrm{E}}$ = $H_{\mathrm{m}}^{\mathrm{E}}$/2 and $G_{\mathrm{m}}^{\mathrm{E}}$ = $H_{\mathrm{m}}^{\mathrm{E}}$. Mixtures with large self-association effects are encountered well above the line $G_{\mathrm{m}}^{\mathrm{E}}$ = $H_{\mathrm{m}}^{\mathrm{E}}$. In view of the $G_{\mathrm{m}}^{\mathrm{E}}/H_{\mathrm{m}}^{\mathrm{E}}$ values given above, one can assume that interactions are mainly of dipolar type. It is pertinent to compare these results with those for the corresponding mixtures with 1-alkanols. In such a case, $G_{\mathrm{m}}^{\mathrm{E}}/H_{\mathrm{m}}^{\mathrm{E}}$(DPE) = 821/741 [90] = 1.11 (1-propanol); 739 [91]/703 [96] = 1.05 (1-butanol). This set of data shows that association effects are more relevant in mixtures with 1-alkanols. On the other hand, DISQUAC predicts, for the 2BuOH + diglyme system, $G_{\mathrm{m}}^{\mathrm{E}}$ = 658 J mol$^{-1}$, and $G_{\mathrm{m}}^{\mathrm{E}}/H_{\mathrm{m}}^{\mathrm{E}}$ = 658/1848 = 0.36, *i.e.*, interactions are essentially dispersive. Note that the values of $TS_{\mathrm{m}}^{\mathrm{E}} = H_{\mathrm{m}}^{\mathrm{E}} - G_{\mathrm{m}}^{\mathrm{E}}$ are positive, *e.g.*, 109 and 331 J mol$^{-1}$ for the 2PrOH or 2BuOH + DPE mixtures, respectively, and 1190 J mol$^{-1}$ for the 2BuOH + diglyme system. Compare with $TS_{\mathrm{m}}^{\mathrm{E}}$(heptane)/J mol$^{-1}$ = 906 [93] − 1183 (DISQUAC result) = −277 (2BuOH); 1621 [97] − 887 (DISQUAC result) = 734 (diglyme).

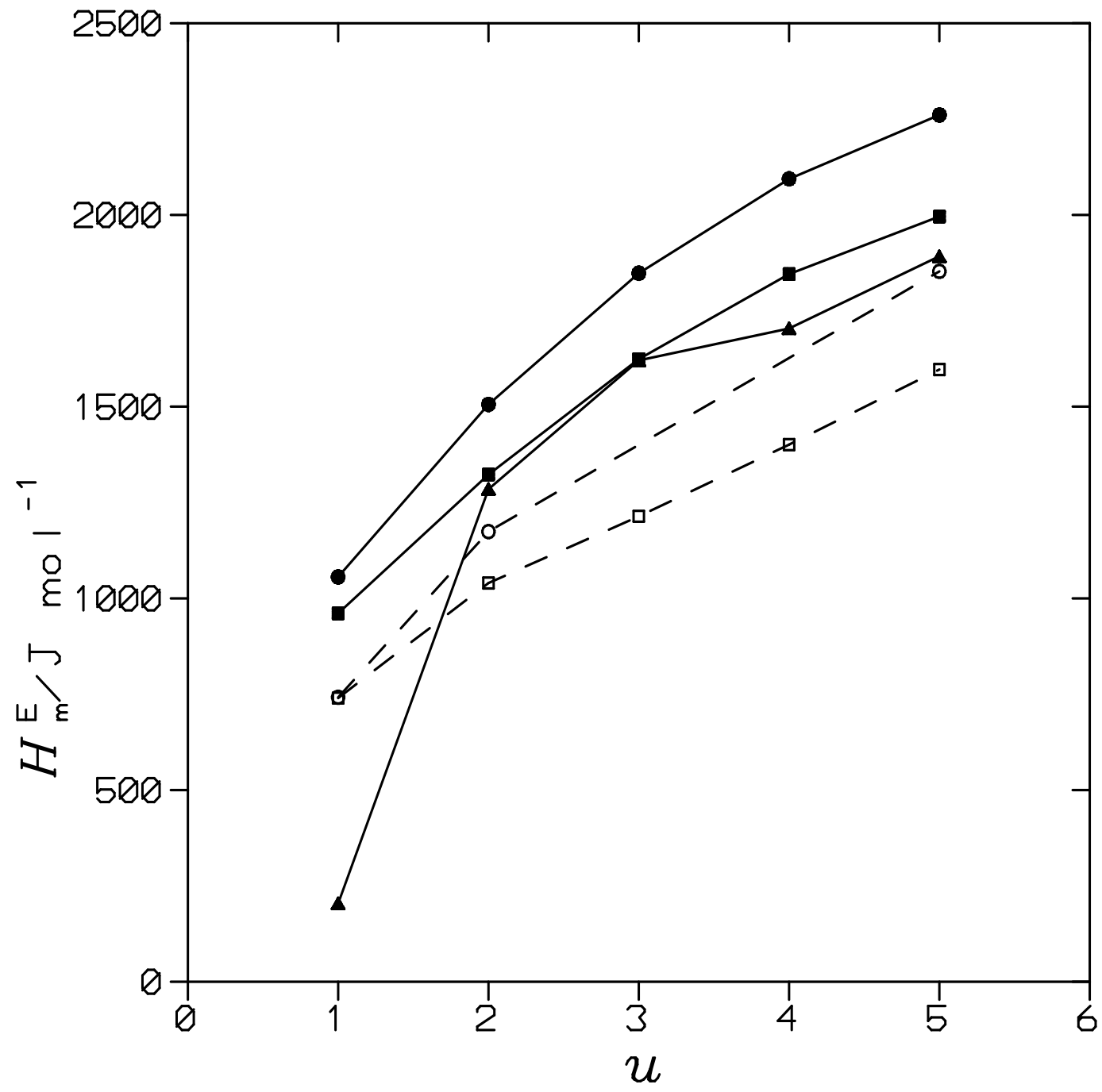

**Figure 6**. $H_m^E$ for several types of glyme mixtures at equimolar composition, temperature $T$ = 298.15 K and pressure $p$ = 95 kPa *vs.* $u$, the number of O atoms in the glyme ($u$ = 1 stands for dipropyl ether). Points: (■), 2-propanol + glyme [28,90]; (●), 2-butanol + glyme ([89] and this work); (▲), glyme + heptane [97–100]; (□), 1-propanol + glyme [90,101–103]; (○), 1-butanol + glyme [96,102,103]. Lines are for the aid of the eye.

For a given linear ether, $H_m^E$(2PrOH) < $H_m^E$(2BuOH) (Figure 6). This may be due to: (i) more interactions between like molecules are broken along the mixing process when 2BuOH is involved. Note that for heptane mixtures, $H_m^E$/J mol$^{-1}$ = 797 (2PrOH) [104] < 906 (2BuOH) [93]. (ii) The number of alcohol-ether interactions created in the mentioned process is higher for solutions with 2PrOH since the OH group is less sterically hindered. (iii) The enthalpies of such interactions are higher (*i.e.*, less negative) for 2BuOH mixtures. Next, we evaluate these enthalpies (termed as $\Delta H_{\text{OH-O}}^{\text{int}}$). If structural effects are neglected [72,105], the $\Delta H_{\text{OH-O}}^{\text{int}}$ values can be evaluated from the expression [6,106]:

$$\begin{aligned}\Delta H_{\text{OH-O}}^{\text{int}} = &\ \bar{H}_{\text{m},1}^{\text{E},\infty}\text{ (2-alkanol + ether)}\\ &-\bar{H}_{\text{m},1}^{\text{E},\infty}\text{(2-alkanol + heptane)}\\ &-\bar{H}_{\text{m},1}^{\text{E},\infty}\text{(ether + heptane)}\end{aligned} \quad (13)$$

where $\bar{H}_{\text{m},1}^{\text{E},\infty}$ stands for the excess partial molar enthalpy at infinite dilution of the first component in the corresponding binary mixture. As in previous applications [27,28], the value of $\bar{H}_{\text{m},1}^{\text{E},\infty}$ (2-alkanol + heptane) is taken equal to 23.2 kJ mol$^{-1}$ and it is considered independent of the alkanol. From our enthalpic data (Table 5), $\bar{H}_{\text{m},1}^{\text{E},\infty}$ (2BuOH)/kJ mol$^{-1}$ = 6.0 ($u$ = 2); 6.3 ($u$ = 3,4); 6.2 ($u$ = 5), and similarly. $\bar{H}_{\text{m},1}^{\text{E},\infty}$ (heptane)/kJ mol$^{-1}$ = 5.5 ($u$ = 2) [98]; 8.2 ($u$ = 3) [97]; 10.8 ($u$ =4) [99]; 13.3 ($u$ = 5) [100]. Consequently, $\Delta H_{\text{OH-O}}^{\text{int}}$/kJ mol$^{-1}$ = −22.7 ($u$ =2); −25.1 ($u$ =3); −27.7 ($u$ = 4); −30.3 ($u$ = 5). These results are close to those of systems with 2PrOH [28]: −23.1 ($u$ = 2); −25.4 ($u$ = 3); −28.5 ($u$ = 4); −30.9 ($u$ = 5), all values in kJ

mol−1. The observed differences between $H_{\mathrm{m}}^{\mathrm{E}}$ values of systems involving a given glyme and 2PrOH or 2BuOH can be then attributed to the points (i) and (ii) stated above. Results for $\Delta H_{\text{OH-O}}^{\text{int}}$ are much higher for systems with DPE, *e.g.*, −16 kJ mol−1 for the 2BuOH solution [27]. Since for a given 2-alkanol, $H_{\mathrm{m}}^{\mathrm{E}}$(glyme) > $H_{\mathrm{m}}^{\mathrm{E}}$(DPE), one can conclude that the number of interactions between like molecules disrupted along mixing is larger when glymes are involved. Note that $TS_{\mathrm{m}}^{\mathrm{E}}$(2BuOH) is much higher for the mixture with diglyme (743 J mol−1) than for the solution with DPE (331 J mol−1).

Finally, it should be mentioned that, for a fixed glyme, $H_{\mathrm{m}}^{\mathrm{E}}$(1-alkanol) < $H_{\mathrm{m}}^{\mathrm{E}}$(isomeric 2-alkanol). The same trend is observed for systems with heptane. Thus, $H_{\mathrm{m}}^{\mathrm{E}}$(heptane)/J mol−1= 581 (1-butanol) [107]; 1056 (2BuOH) [93]. The values of $\Delta H_{\text{OH-O}}^{\text{int}}$ for mixtures with 1-propanol or 1-butanol are quite close to those reported above for 2-alkanol systems. For example, $\Delta H_{\text{OH-O}}^{\text{int}}(u = 2)$/kJ mol−1 = −23.9 (1-propanol); −23.4 (1-butanol), or $\Delta H_{\text{OH-O}}^{\text{int}}(u = 5)$ /kJ mol−1 = −31.9 (1-propanol); −31 (1-butanol) [7]. Therefore, the higher $H_{\mathrm{m}}^{\mathrm{E}}$ results obtained when 1-alkanol is replaced by an isomeric 2-alkanol can be explained in similar terms to the previous ones (see above).

## 7.2. Volumetric data

$V_{\mathrm{m}}^{\mathrm{E}}$ is the result of several contributions: (i) the breaking of interactions between like molecules contribute positively to this property; (ii) interactions between unlike molecules and structural effects contribute negatively to it. For 2BuOH + glyme systems, both $V_{\mathrm{m}}^{\mathrm{E}}$ and $H_{\mathrm{m}}^{\mathrm{E}}$ are positive and this means that the main contribution to $V_{\mathrm{m}}^{\mathrm{E}}$ arises from interactional effects. According to this statement, the values of $A_p$ are positive (Table 3, Figure S3). However, $V_{\mathrm{m}}^{\mathrm{E}}$ increases when 2,5-dioxahexane is replaced by 2,5,8-trioxanonane and then decreases (Table 10, Figure 7). Such a decrease may be ascribed to increased structural effects, particularly related to the large difference of the molar volumes of the mixture compounds. Note that the mentioned difference increases with $u$ (Table S1). In the case of 2PrOH mixtures, $V_{\mathrm{m}}^{\mathrm{E}}$ varies differently: its magnitude changes from a positive value for $u$ = 2 to a slightly negative result for $u$ = 5, (Table 10, Figure 7), which can be explained in similar terms to those used above. In the case of systems with DPE, the negative $V_{\mathrm{m}}^{\mathrm{E}}$ values (Table 10, Figure 7) indicate that the contribution from structural effects to this excess function is dominant. The effects mentioned are of free volume type (note that the $\alpha_{pi}$ values differ largely, Table S1) and are also due to the difference in size of the solution compounds.

For a fixed linear polyether, $V_{\mathrm{m}}^{\mathrm{E}}$(1-alkanol) < $V_{\mathrm{m}}^{\mathrm{E}}$(isomeric 2-alkanol) (Figure 7), in agreement with the corresponding $H_{\mathrm{m}}^{\mathrm{E}}$ variation. Nevertheless, it should be mentioned that for a number of (1-alkanol + glyme) mixtures, the $V_{\mathrm{m}}^{\mathrm{E}}$ values are negative [49,101,108], and, in such cases, $V_{\mathrm{m}}^{\mathrm{E}}$ is determined mainly by structural effects. However, the values of $A_p$ can be positive or negative. Positive values are encountered for (1-propanol + glyme) systems [108] (1.23 $10^{-3}$ cm3 mol-1 K-1 for the monoglyme solution), which means that the interactional contribution to $V_{\mathrm{m}}^{\mathrm{E}}$ and $T$ increase in line. In contrast, the 1-octanol + monoglyme mixture shows a negative $A_p$ value (−1.72 $10^{-3}$ cm3 mol−1 K−1 [49]), and, in this case, the structural contribution to $V_{\mathrm{m}}^{\mathrm{E}}$ and $T$ increase in line.

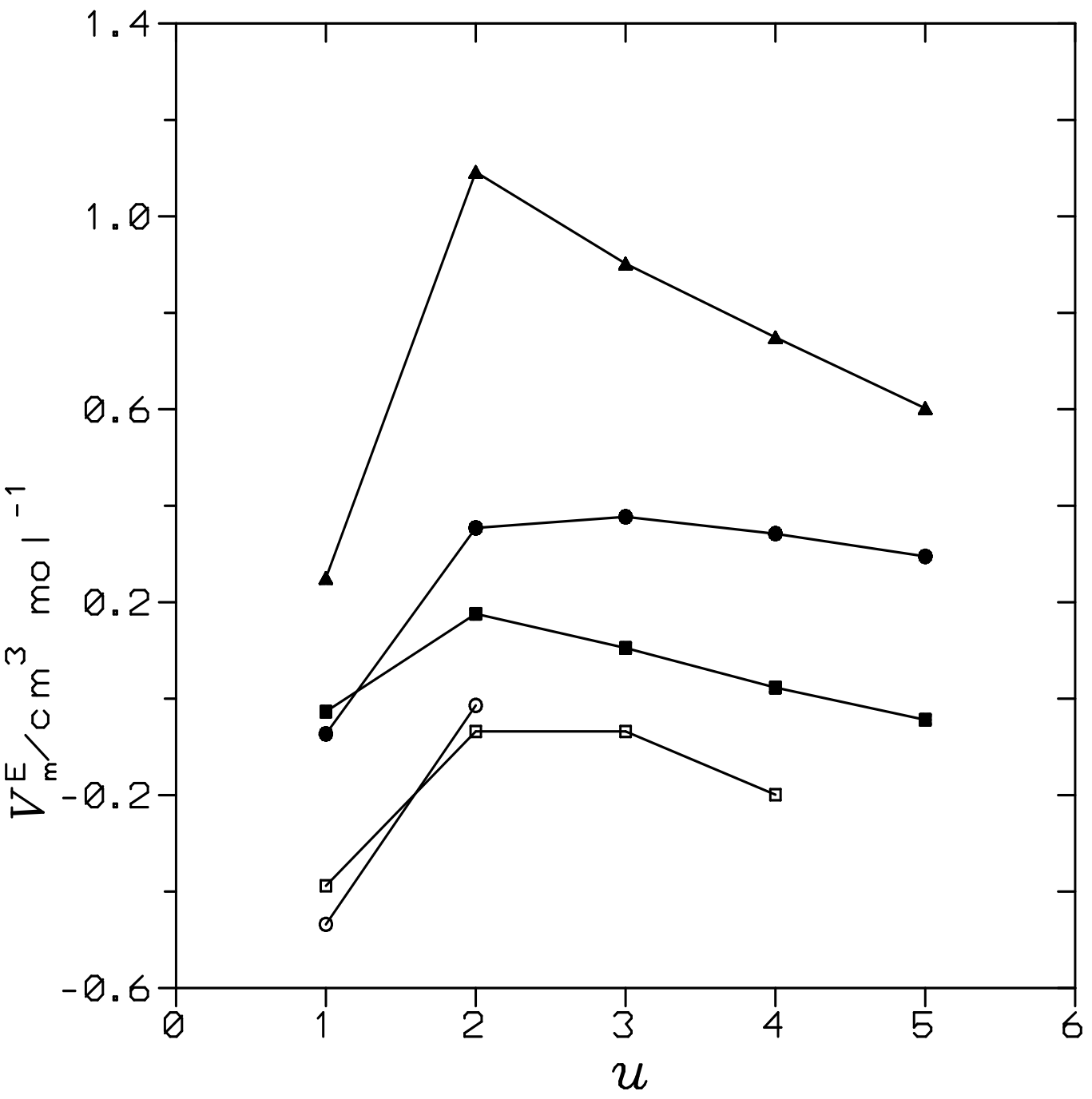

**Figure 7**. $V_m^E$ for several types of glyme mixtures at equimolar composition, temperature $T$ = 298.15 K and pressure $p$ = 95 kPa *vs.* $u$, the number of O atoms in the glyme ($u$ = 1 stands for dipropyl ether). Points: (■), 2-propanol + glyme [28,90]; (●), 2-butanol + glyme ([89] and this work); (▲), glyme + heptane [64,109–111]; (□), 1-propanol + glyme [90,101,108,112]; (○), 1-butanol + glyme [113,114]. Lines are for the aid of the eye.

## 7.3. Theoretical results

### *7.3.1. DISQUAC*

As already mentioned, the model describes well the thermodynamic properties of the systems under study. It is remarkable that DISQUAC predicts correctly the decrease of $G_m^E$ when $T$ is increased (Table 9). For glyme systems, interesting results arise from the dependence of the interchange coefficients with the molecular structure (Table 6). (i) The $C_{eh,l}^{QUAC}$ ($l$ = 1,2) coefficients are practically independent on the mixture compounds. This is a common behavior in the framework of DISQUAC, since often the QUAC parameters remain constant along a given homologous series [115–117]. (ii) We note that the values of $C_{eh,2}^{QUAC}$ are much lower than those of $C_{eh,2}^{DIS}$ and this suggests that the interactions related to the (e,h) contacts are mainly of dispersive type. On the other hand, the values of $C_{eh,2}^{QUAC}$ for the 2-alkanol + glyme systems are considerably lower than those for DPE solutions, which remarks that orientational effects linked to the (e,h) contact are stronger in the latter mixtures.

### *7.3.2. UNIFAC (Dortmund)*

The model provides very poor results for 2-propanol + linear ether mixtures. This can be explained as follows: (i) the same interaction parameters are used for solutions including primary, secondary and tertiary alcohols. These alkanols merely differ by the geometrical parameters of the subgroups within the same main group (No. 5) [30]. In the framework of DISQUAC, the geometrical parameters of the OH group are independent of the alcohol, but mixtures with primary, secondary or tertiary alkanols are characterized by different interaction parameters [81,88,118]; (ii) the same UNIFAC interaction parameters are used for mixtures containing any type of linear ether, which ignores, *e.g.*, the strong proximity effects present in glyme + *n*-alkane systems [17,80]. For example, UNIFAC predicts $H_m^E$(heptane)/J mol$^{-1}$ = 814 ($u$ = 2); and 945 ($u$ = 5). The experimental results (in the same units) are: 1621 [98] and 1897 [100], respectively.

### 7.3.3. ERAS

The model fails when describing the experimental $H_m^E$ results for 2-alkanol + glyme systems (Table 8, Figure 3), while describes correctly the corresponding values for mixtures with DPE (Table 8, Figure S4). This clearly indicates that, in the former solutions, association effects are of minor importance and that physical interactions are dominant. Note that the Flory model provides excellent $H_m^E$ results for the mixtures 1-alkanol (other than methanol or ethanol) + linear polyether [7] or 2PrOH + glyme [28]. Better results have been attained for 1-alkanol + linear polyether systems assuming that the ethers are self-associated compounds [101], which does not seem to be very pertinent. It is remarkable that ERAS provides fairly good results for $V_m^E$ of the systems under consideration (Table 10, Figure 5).

Some comments on the ERAS parameters can be provided (Table 7). (i) $K_{AB}$ increases along a certain homologous series (2-alkanol fixed), indicating that more interaction between unlike molecules is created upon mixing, probably due to the increasing number of O atoms in the ether. On the other hand, for a given oxaalkane, $K_{AB}$(2BuOH) < $K_{AB}$(2PrOH), which can be explained because the OH group is more sterically hindered in 2BuOH. (ii) The values of $X_{AB}$ reveal that physical interactions are much less relevant in systems with DPE, and ERAS describes correctly the corresponding $H_m^E$ curves. Note that, in this case, the $\Delta h_{AB}^*$ parameter is not far from the $\Delta H_{OH\text{-}O}^{int}$ value (see above). (iii) The large negative value of the $\Delta v_{AB}^*$ parameter for the 2BuOH + DPE system is noticeable. It merely remarks that, in order to describe the large structural effects present in the solution, the positive physical contribution to $V_m^E$ must be overcompensated by a larger negative contribution linked to association effects.

# 8. Conclusions

$H_m^E$ and $V_m^E$ data have been reported for 2BuOH + glyme mixtures. Results indicate that interactions between like molecules are dominant by far, and that interactions are mainly dispersive. These features largely overcompensate the strong alkanol-ether interactions. The interactional contribution to $V_m^E$ is higher than that from structural effects. 2-Alkanol + DPE, or + glyme systems have been studied using the DISQUAC, UNIFAC (Dortmund) and ERAS models. ERAS results underline that glyme systems are characterized by physical interactions, and that association effects are much more relevant in mixtures with DPE. The DISQUAC interaction parameters suggest that interactions linked to the (e,h) contacts are essentially dispersive. UNIFAC represents poorly the thermodynamic properties of the studied solutions.

# Funding

This work was supported by Project PID2022-137104NA-I00 funded by MICIU/AEI/10.13039/501100011033/ and by FEDER, UE, and also by Project VA168P25 funded by Junta de Castilla y León and European Regional Development Fund (FEDER, UE). F. Pazoki acknowledges the FPI grant PREP2022-000047 from MCIN/AEI/10.13039/501100011033/ and FEDER, UE. The author J. V. Alves-Laurentino would like to acknowledge funding from European Social Fund Plus, Operational Programme of Castilla y León, and from Junta de Castilla y León, through Consejería de Educación.

# CRediT author statement

**Fatemeh Pazoki**: Formal analysis, Investigation, Visualization, Writing - Review & Editing. **João Victor Alves-Laurentino**: Software, Data Curation, Visualization. **Susana Villa**: Validation, Formal analysis, Writing - Review & Editing. **Daniel Lozano-Martín**: Investigation, Writing - Review & Editing, Supervision, Project administration, Funding acquisition. **Fernando Hevia**: Writing - Original Draft, Writing - Review & Editing, Supervision, Project administration, Funding acquisition. **Luis Felipe Sanz**: Methodology, Software, Resources. **Juan Antonio González**: Conceptualization, Methodology, Formal analysis, Visualization, Writing - Original Draft.

# Density and excess molar enthalpy of 2-butanol + glyme liquid mixtures. Application of the DISQUAC, UNIFAC (Dortmund) and ERAS models to 2-alkanol + linear ether systems

## Supplementary material

Fatemeh Pazoki, João Victor Alves-Laurentino, Susana Villa, Daniel Lozano-Martín, Fernando Hevia*, Luis Felipe Sanz, Juan Antonio González

GETEF. Departamento de Física Aplicada. Facultad de Ciencias. Universidad de Valladolid. Paseo de Belén, 7, 47011 Valladolid, Spain.

* Corresponding author, e-mail: luisfernando.hevia@uva.es

# Tables and figures

**Table S1**. Physical properties and ERAS parameters[a] at 298.15 K for pure compounds used in this work.

| Compound[b] | $V_{\mathrm{m}i}$/ cm$^3$ mol$^{-1}$ | $\alpha_{pi}$/ $10^{-3}$ K$^{-1}$ | $\kappa_{Ti}$/ $10^{-12}$ Pa$^{-1}$ | $V^*_{\mathrm{m}i}$/ cm$^3$mol$^{-1}$ | $P^*_i$/ J cm$^{-3}$ | $K_i$ | $\Delta h^*_i$/ kJ mol$^{-1}$ | $\Delta v^*_i$/ cm$^3$ mol$^{-1}$ |
|---|---|---|---|---|---|---|---|---|
| 2PrOH | 76.96 [S1] | 1.082 [S1] | 1110 [S2] | 62.09 | 398.8 | 131 [S3] | −25.1 [S3] | −5.6 [S3] |
| 2BuOH | 92.39[c] | 1.036[c] | 970 [S4] | 75.30 | 425.0 | 68 [S5] | −25.1 [S5] | −5.6 [S5] |
| monoglyme | 104.58 [S1] | 1.272 [S1] | 1114 [S6] | 80.38 | 575.9 | 0 | 0 | 0 |
| diglyme | 142.93 [S1] | 1.059 [S1] | 822 [S6] | 113.47 | 609.8 | 0 | 0 | 0 |
| triglyme | 181.84 [S1] | 0.969 [S1] | 707 [S6] | 146.49 | 629.6 | 0 | 0 | 0 |
| tetraglyme | 220.89 [S1] | 0.920 [S1] | 644 [S7] | 179.43 | 645.5 | 0 | 0 | 0 |
| DPE | 137.68 [S5] | 1.261 [S5] | 1440 [S5] | 106.00 | 440.5 | 0 | 0 | 0 |

[a] $V_{\mathrm{m}i}$, molar volume; $\alpha_{pi}$, isobaric thermal expansion coefficient; $\kappa_{Ti}$, isothermal compressibility; $V^*_{\mathrm{m}i}$ and $P^*_i$, reduction parameters for volume and pressure, respectively; $K_i$, equilibrium constant; $\Delta h^*_i$, molar enthalpies of intermolecular hydrogen-bonding; $\Delta v^*_i$, molar hydrogen-bonding volumes.

[b] Symbols: 2PrOH, 2-propanol; 2BuOH, 2-butanol; DPE, dipropyl ether.

[c] This work.

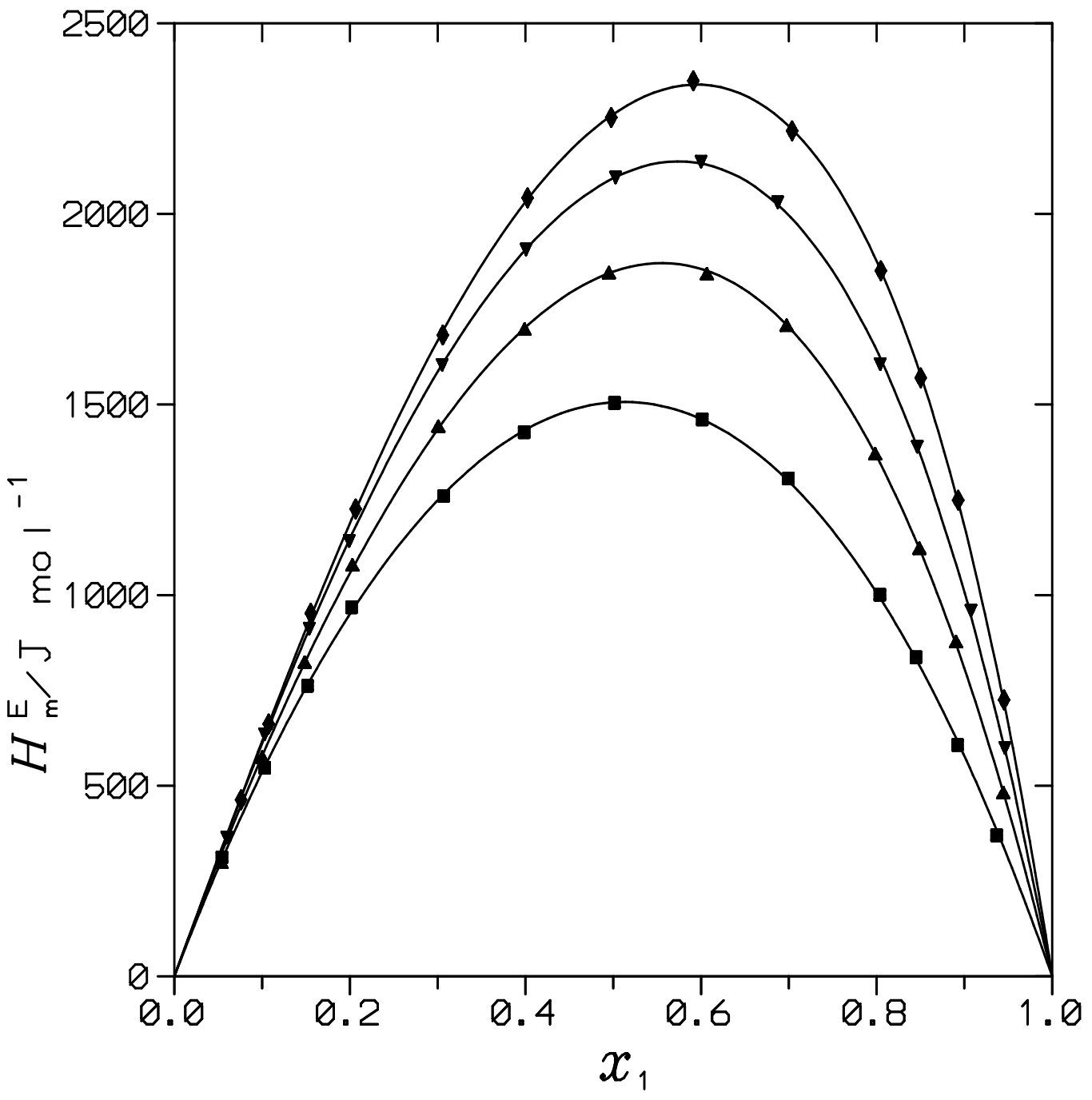

**Figure S1**. $H^E_m$ for 2-butanol (1) + linear polyether (2) mixtures at 298.15 K and 95 kPa. Points, experimental results (this work): (■), 2,5-dioxahexane; (▲), 2,5,8-trioxanonane; (▼), 2,5,8,11-tetraoxadodecane; (◆), 2,5,8,11,14-pentaoxapentadecane. Solid lines, results from Redlich-Kister expansions using parameters listed in Table 5.

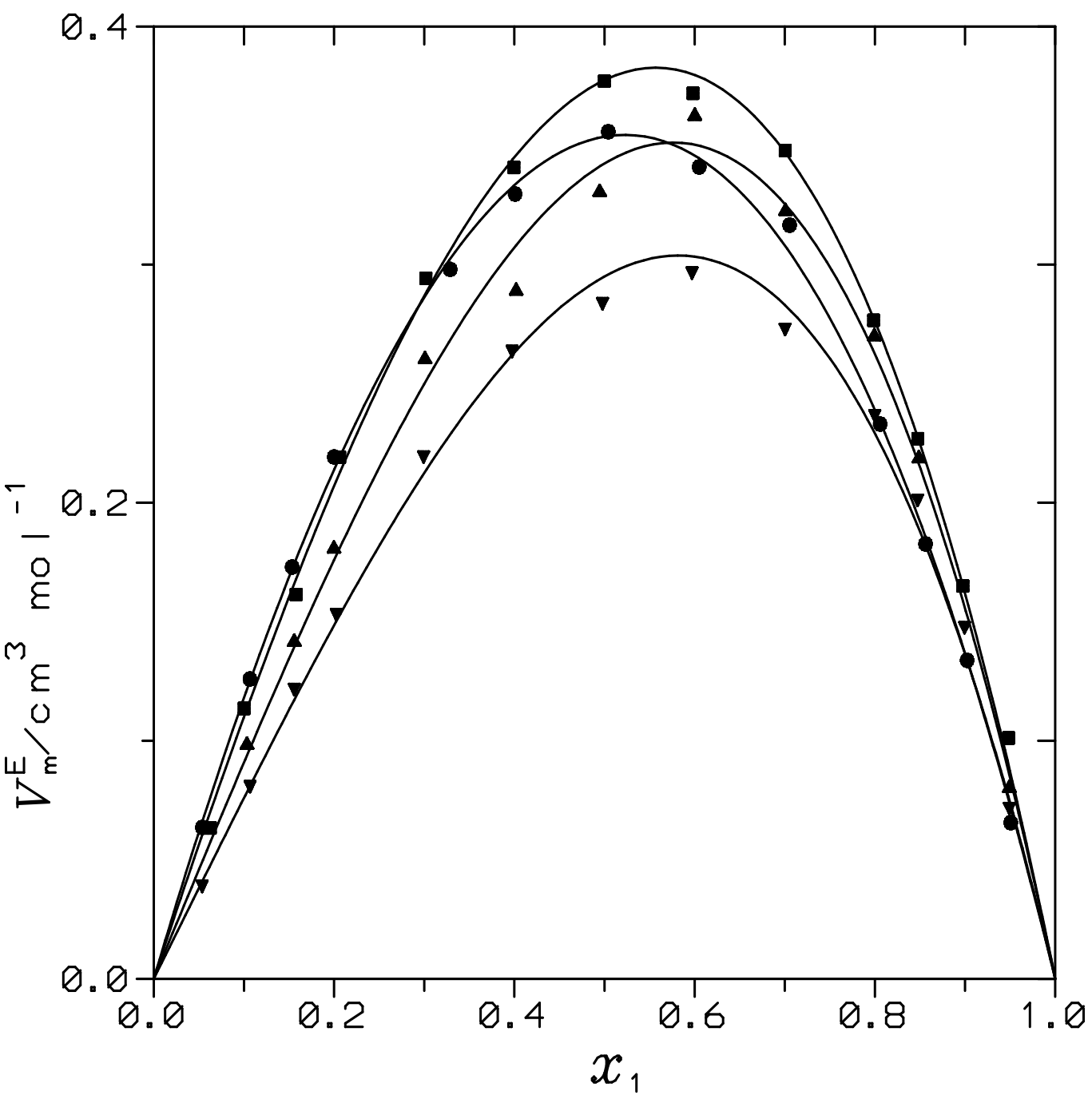

**Figure S2**. $V^E_m$ for 2-butanol (1) + linear polyether (2) mixtures at 298.15 K and 95 kPa. Points, experimental results (this work): (●), 2,5-dioxahexane; (■), 2,5,8-trioxanonane; (▲), 2,5,8,11-tetraoxadodecane; (▼), 2,5,8,11,14-pentaoxapentadecane. Solid lines, results from Redlich-Kister expansions using parameters listed in Table 5.

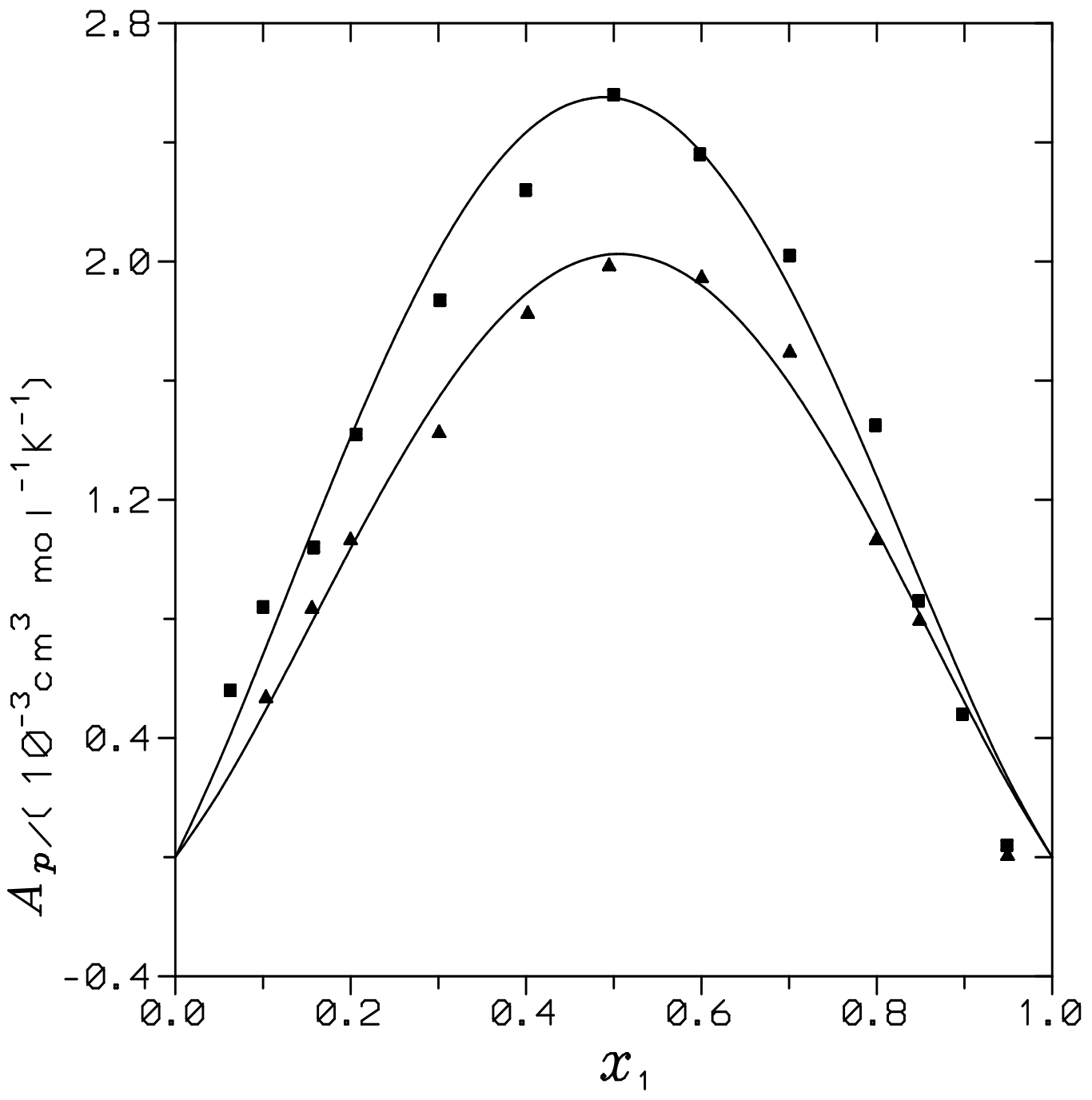

**Figure S3**. $A_p = (\partial V_m^E/\partial T)_{p,x}$ for 2-butanol + glyme systems at 298.15 K and 95 kPa. Points, experimental results: (▲), 2,5,8-trioxanonane; (■), 2,5,8,11,14-pentaoxapentadecane. Solid lines, results from Redlich-Kister expansions using parameters listed in Table 5.

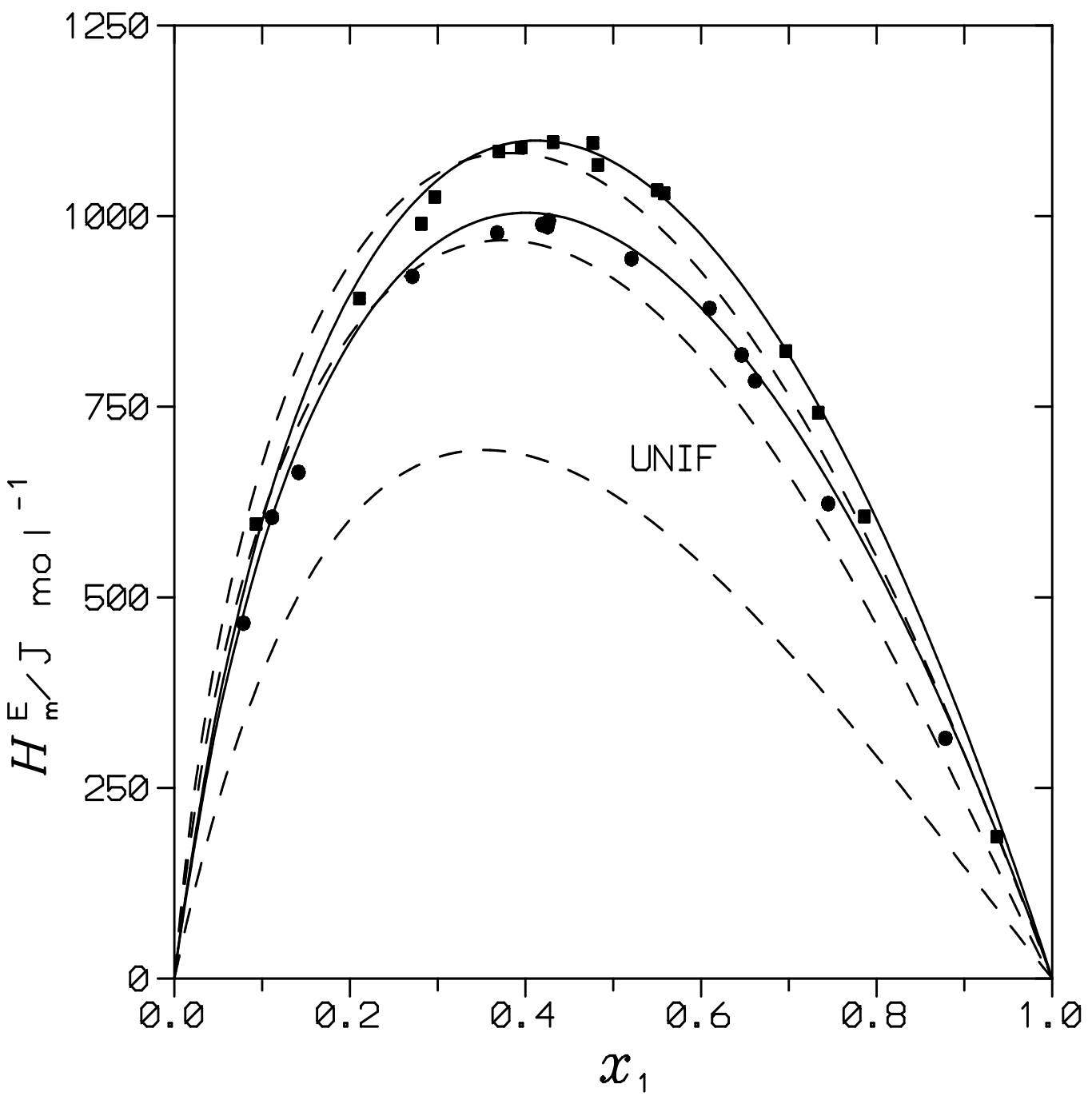

**Figure S4**. $H_m^E$ for 2-propanol (1), or 2-butanol (1) + dipropyl ether (2) mixtures at 298.15 K and atmospheric pressure. Points, experimental results: (●), 2-propanol [S8]; (■), 2-butanol [S5]. Solid lines, DISQUAC calculations using interchange coefficients listed in Table 6. Dashed lines, ERAS results using parameters listed in Table 7, or UNIFAC values for the 2-propanol mixture with interaction parameters from the literature [S9].